\documentclass[twocolumn, times, tighten,twocolappendix]{aastex63}

\usepackage{graphicx,multirow,hyperref,url,color,xspace}

\newcommand{\msun}{{\rm M}_{\sun}}
\newcommand{\ledd}{L_{{\rm Edd}}}
\newcommand{\rg}{{R_{\rm{g}}}}
\newcommand{\xmm}{{\textit{XMM-Newton}}\xspace}
\newcommand{\xte}{{\textit{RXTE}}\xspace}

\newcommand{\swift}{{\textit{Neil Gehrels Swift}}\xspace}

\newcommand{\source}{{XTE J1752--223}\xspace}

\begin{document}

\defcitealias{Garcia18}{G18}
\newcommand{\garcia}{{\citetalias{Garcia18}}\xspace}

\defcitealias{Munoz10}{MD10}
\newcommand{\munoz}{{\citetalias{Munoz10}}\xspace}

\title{Does the disk in the hard state of XTE J1752--223 extend to the innermost stable circular orbit?}
\shorttitle{The disk in the hard state of XTE J1752--223}

\author{Andrzej A. Zdziarski}
\affil{Nicolaus Copernicus Astronomical Center, Polish Academy of Sciences, Bartycka 18, PL-00-716 Warszawa, Poland; \href{mailto:aaz@camk.edu.pl}{aaz@camk.edu.pl}}
\author{Barbara De Marco}
\affil{Departament de F{\'{\i}}sica, EEBE, Universitat Polit{\`e}cnica de Catalunya, Av.\ Eduard Maristany 16, E-08019 Barcelona, Spain; \href{mailto:barbara.de.marco@upc.edu}{barbara.de.marco@upc.edu}}
\author{Micha{\l} Szanecki}
\affil{Nicolaus Copernicus Astronomical Center, Polish Academy of Sciences, Bartycka 18, PL-00-716 Warszawa, Poland; \href{mailto:aaz@camk.edu.pl}{aaz@camk.edu.pl}}
\author{Andrzej Nied{\'z}wiecki}
\affil{Faculty of Physics and Applied Informatics, {\L}{\'o}d{\'z} University, Pomorska 149/153, PL-90-236 {\L}{\'o}d{\'z}, Poland; \href{mailto:andrzej.niedzwiecki@uni.lodz.pl}{andrzej.niedzwiecki@uni.lodz.pl}}
\author{Alex Markowitz}
\affil{Nicolaus Copernicus Astronomical Center, Polish Academy of Sciences, Bartycka 18, PL-00-716 Warszawa, Poland; \href{mailto:aaz@camk.edu.pl}{aaz@camk.edu.pl}}

\shortauthors{Zdziarski et al.}

\begin{abstract}
The accreting black-hole binary XTE J1752--223 was observed in a stable hard state for 25\,d by RXTE, yielding a 3-140\,keV spectrum of unprecedented statistical quality. Its published model required a single Comptonization spectrum reflecting from a disk close to the innermost stable circular orbit. We studied that model as well as a number of other single-Comptonization models (yielding similarly low inner radii), but found they violate a number of basic physical constraints, e.g., their compactness is much above the maximum allowed by pair equilibrium. We also studied the contemporaneous 0.55--6\,keV spectrum from the Swift/XRT and found it well fitted by an absorbed power law and a disk blackbody with the innermost temperature of 0.1\,keV. The normalization of the disk blackbody corresponds to an inner radius of $\gtrsim$20 gravitational radii and its temperature, to irradiation of the truncated disk by a hot inner flow. We have also developed a Comptonization/reflection model including the disk irradiation and intrinsic dissipation, but found that it does not yield any satisfactory fits. On the other hand, we found that the $\leq$10\,keV band from RXTE is much better fitted by a reflection from a disk with the inner radius $\gtrsim$100 gravitational radii, which model then underpredicts the spectrum at $>$10\,keV by $<$10\%. We argue that the most plausible explanation of the above results is inhomogeneity of the source, with the local spectra hardening with the decreasing radius. Our results support the presence of a complex Comptonization region and a large disk truncation radius in this source.
\end{abstract}

\section{Introduction}
\label{intro}

Broad-band, $\sim$1--300\,keV, X-ray spectra of black hole (BH) binaries in the (low) hard state (LHS) are often successfully fitted by the sum of three components: blackbody emission from a disk, Comptonization in a hot plasma (sometimes approximated by a power law with an exponential high-energy cutoff), and a reflected spectrum from the plasma irradiating the disk (e.g., \citealt{Dove97b, Zdziarski98, DGK07} and references therein). When data at $\lesssim$3\,keV are not available, the disk component is usually not detectable, and the spectra can be well fitted by Comptonization and reflection alone (e.g., \citealt{Garcia15,Dzielak19}). This is, in particular, the case for the \xte spectrum from the BH X-ray binary \source studied by \citet{Garcia18}, hereafter \citetalias{Garcia18}. 

Still, the usual exposure time used to make an X-ray spectrum is several thousands of seconds (and often more), which is much longer than the dominant variability time scale in the LHS of the order of a second (corresponding to the peak of the variability power per logarithm of frequency). The total root-mean-square (rms) variability in the LHS is typically $\sim$30--40\% \citep{Munoz11}. This means that the fitted X-ray spectra are averaged over many cycles of strong variability. It could be a spectrum of a constant shape varying in its amplitude but rms spectra in this state often decrease with the energy in the X-ray regime, thus inducing significant spectral variability, see, e.g., \citet{GZ05,Stiele15}. This could still correspond to a single Comptonizing region with a variable input of soft seed photons (presumably from the disk), resulting in a pivoting spectrum \citep{ZPP02,GZ05}. 

Further constraints on the structure of the sources are obtained from combined spectral/timing analysis. \citet{RGC99} showed that X-ray spectra corresponding to different Fourier-frequency ranges of the variability of Cyg X-1 in its LHS become harder with the increasing frequency. \citet{Axelsson18} performed similar calculations, but fitted the power spectrum with three Lorentzian components, and obtained the X-ray spectra corresponding to each. Similarly, they found that the X-ray spectra harden with the increasing Lorentzian peak frequency. This technique was later used by \citet{Mahmoud18b} and \citet{Mahmoud19} to develop detailed models of accretion flows fitting simultaneously observational constraints from the spectra, variability and time lags, and showing that there have to be more than one Comptonization component, which also agrees with the spectral/timing analysis of \citet{Yamada13}. In this framework, the hard X-ray lags are interpreted as due to the local spectra becoming harder \citep{Kotov01}. Therefore, even if a good spectral fit can be found to a time-averaged X-ray spectrum with a single-Comptonization and reflection model, such a model appears to be not physically consistent with combined spectral/timing results. 

Here, we consider the case of the LHS of the X-ray binary \source. It is a transient accreting BH binary, and it has had only one outburst so far, in 2009--2010 \citep{Shaposhnikov10}. The BH nature of the system is inferred from the overall similarity of its properties to known BH systems. The distance, $d$, and binary parameters of the source remain relatively uncertain. \citet{Shaposhnikov10} estimated $d\approx 3.5\pm 0.4$\,kpc and the BH mass of $M\approx 9.6\pm 0.9\msun$ using a correlation method. (We note, however, that the same method yielded $M=7.9\pm 1.0\msun$ for Cyg X-1, much below the current dynamical estimate.) Then, \citet{Ratti12} estimated $d\approx 9.1\pm 4.5$\,kpc based on the flux of the soft-to-hard transition, and $3.5\,{\rm kpc}\lesssim d\lesssim 8$\,kpc including also constraints on the emission of the donor, with some preference for $\sim$8\,kpc based on the radio/X-ray correlation and consideration of the orbital period. A rough constraint on $d^2/M$ can also be set from the remarkable similarity of the X-ray hardness-count rate diagram for \source to that of GX 339--4, shown in fig.\ 1 of \garcia. The distance to GX 339--4 was estimated in \citet{Zdziarski19} as 8--12\,kpc, approximately correlated with the BH mass estimate within 4--$11\msun$, which favors a large distance for \source. That hardness-count rate shape is also characteristic of low-inclination BH transients \citep{Munoz13}, which implies $i\lesssim 60\degr$. The binary inclination is then limited to $i\lesssim 80\degr$ \citep{Ratti12}, while that of the jet to $i\lesssim 49\degr$ \citep{Miller_Jones11}.  

A remarkable feature of \source is that, during the rise, it was observed for 25\,d by \xte in a stable LHS, with an almost constant flux and hardness \citepalias{Garcia18}. Still, the source was highly variable on short time scales, with a $\approx 48\%$ rms variability in the $E=2$--15 keV energy range and the 0.002--128 Hz frequency range measured during the first 4\,d of the \xte observations (\citealt{Munoz10}; hereafter \munoz). \citetalias{Garcia18} has obtained the average spectra from the PCA (3--45\,keV) and HEXTE (20--140\,keV) for all of these observations. This resulted in highly accurate spectra with a 300 ks exposure and very large count numbers, $\approx\! 10^8$ and $\approx\! 10^7$ for the PCA and HEXTE, respectively. In spite of the relatively low energy resolution of the PCA, the accuracy of the calibration and the very high count numbers allow for detailed spectral fits, as shown by \citetalias{Garcia18}. 

\citetalias{Garcia18} fitted thermal Comptonization and relativistically broadened reflection using the {\tt xillver} and {\tt relxill} models \citep{Garcia13,Dauser16}. They found extreme relativistic broadening, with their overall best fits corresponding to the disk inner radius, $R_{\rm in}$, very close to the radius of the innermost stable circular orbit (ISCO), $R_{\rm ISCO}$, and the primary source forming a `lamppost' (a static point source on the BH rotation axis) very close to the BH horizon radius, $R_{\rm hor}$. This result can be considered to be one of the best existing arguments for the disk reaching the immediate vicinity of the ISCO in the luminous LHS.

The question whether the disk in accreting BH binaries reaches the ISCO already in the luminous LHS (at $\gtrsim 0.01$ of the Eddington luminosity, $\ledd$) or only in the soft state has, however, been the subject of an intense ongoing debate. Results similar to those of \citetalias{Garcia18} have been obtained in many other studies (e.g., \citealt{Reis08, Reis10, Garcia15,Garcia19}). On the other hand, spectral fits obtaining significantly truncated disks in the luminous LHS have been obtained as well (e.g., \citealt{Plant15,Basak16,Basak17,Dzielak19}). Significant truncation is also implied by the measured relatively long reverberation lags of soft X-rays responding to variable hard X-rays \citep{DeMarco15,DeMarco17,Mahmoud19}, modelling of type C QPOs as precession of the inner hot disk (e.g., \citealt{Ingram16}), and accounting for the re-emission of the X-rays absorbed by the disk \citep{ZDM20}.

Given the above arguments in favor of an inhomogeneous accretion flow in the LHS of BHXRBs, and the X-ray timing characteristics of the source (\munoz), here we reconsider the \xte data from \source, with the aim of testing this scenario. Crucially, we also study data from the X-ray Telescope (XRT; \citealt{Burrows00}) onboard \swift. The XRT performed 10 short observations of \source during the first 9 days of the \xte observations. These data, not considered in \garcia, extend the spectral coverage down to $\approx$0.5\,keV, allowing us to constrain the presence of a disk blackbody component.

\section{Observations}
\label{observations}

Details about the \xte observations and their data reduction are given in \garcia. They combined the 57 individual pointings taken during MJD 55130--55155 using the method of \citet{Garcia15}, which accounts for changes in the hardness ratio, slightly decreasing during these observations, see fig.\ 1 in \garcia. Then they applied the calibration corrections of \citet{Garcia14,Garcia16b} to the resulting average spectra from the PCA and HEXTE, respectively, aimed at improving the quality of the spectra. We use the same combined and corrected PCA and HEXTE spectral data as \garcia. The PCA data include a 0.1\% systematic error, and none is added to the HEXTE data.

We analyzed the XRT observations carried out during MJD 55130--55138 in the Windowed Timing mode, with the total exposure of 9376\,s. The data were reduced following standard procedures ({\tt xrtpipeline} within HEASOFT v.6.25). Source counts were extracted from a circular region with radius of $47''$ centered on the source. The background counts were extracted from an annular region centered on the source. We selected only single-pixel events ({\tt grade}=0).  Ancillary response files were generated using the {\tt xrtmkarf} task, and the response file {\tt swxwt0s6$\_$20090101v015.rmf} was used. The spectra were summed into a single spectrum using the {\sc ftool} {\tt addspec}, which also creates an appropriate ancillary response file for the summed spectrum. The spectrum has been rebinned requiring the minimum number of channels per bin of 3 and a signal to noise ratio of 50, and a 1\% systematic error has been added. We have also looked for a possible contribution of a scattering halo, but found it negligible.

\section{Fits to the X-ray Spectra of \source}
\label{fits}

We study the spectra using the X-ray fitting package {\sc{xspec}} \citep{Arnaud96}. The reported fit uncertainties are for 90\% confidence, $\Delta\chi^2 \approx 2.71$. 

As in \garcia, the overall slope of the PCA data is corrected using the model {\tt jscrab} \citep{Steiner10}, which multiplies the spectrum by a power law with a residual index difference, $\Delta\Gamma=0.01$ (where $\Gamma$ is defined by the photon number flux $\propto E^{-\Gamma}$) and a normalization factor of 1.097, to achieve an agreement with the standard Crab results, see \garcia for details. The same model is applied to the HEXTE and XRT data with free $\Delta\Gamma$ and the relative normalization. This allows to partially correct for differences between the spectral calibration of the three instruments. As found in \garcia, the PCA data still suffer from some calibration problems at highest detector energies, which are corrected by adding two narrow Gaussian lines at 29.8 and 43.4\,keV. Similarly, as found by \munoz, fitting the $\sim$1.5--2.5\,keV XRT data requires an addition of two instrumental lines, at $\approx$1.7 and $\approx$2.3\,keV. These lines have negligible effect on the fit parameters. 

We account for the ISM absorption using the {\tt tbabs} model \citep{WAMC00}. We have found that both the fitted value of the H column density as well the actual absorption of low-energy parts of the spectra depend sensitively on the assumed abundances. The fitted absorption is much stronger for the abundances of \citet{WAMC00} compared to those of \citet{AG89}. This is especially the case for the XRT data, whose fitting yields unrealistically strong disk blackbody components when using the former abundances. Also, the total H{\sc i} and H$_2$ column density in the direction of \source is\footnote{\url{https://www.swift.ac.uk/analysis/nhtot/index.php}} $N_{\rm H}\approx 6.0\times 10^{21}$\,cm$^{-2}$ \citep{Willingale13}, whereas fitting with the former abundances yields almost twice that value. Therefore, we use here the abundances of \citet{AG89}, except when noted otherwise.

\subsection{The XRT spectrum}
\label{S:XRT}

\begin{figure}
  \centering
  \includegraphics[height=7.5cm,angle=-90]{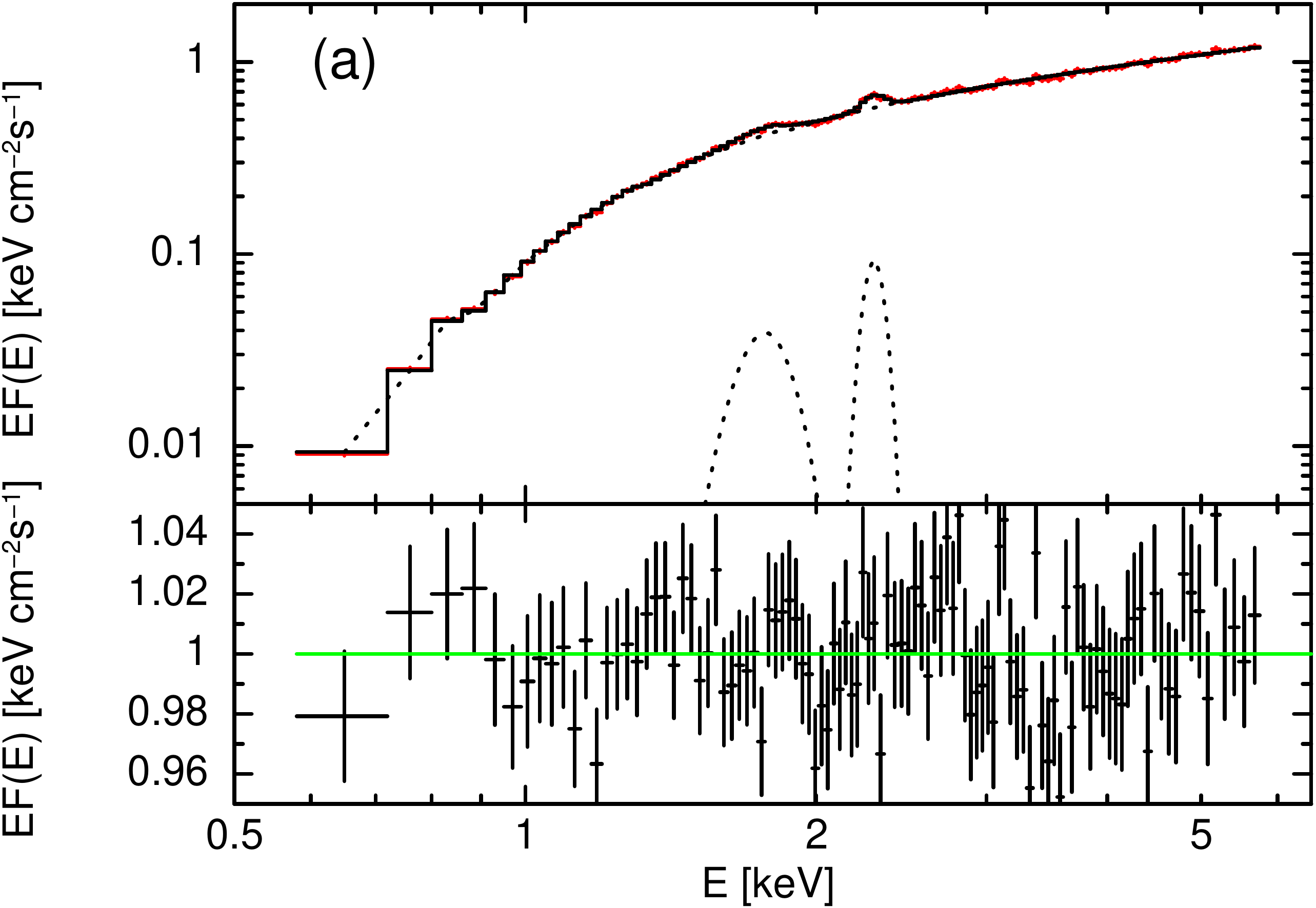}
  \includegraphics[height=6.cm,angle=-90]{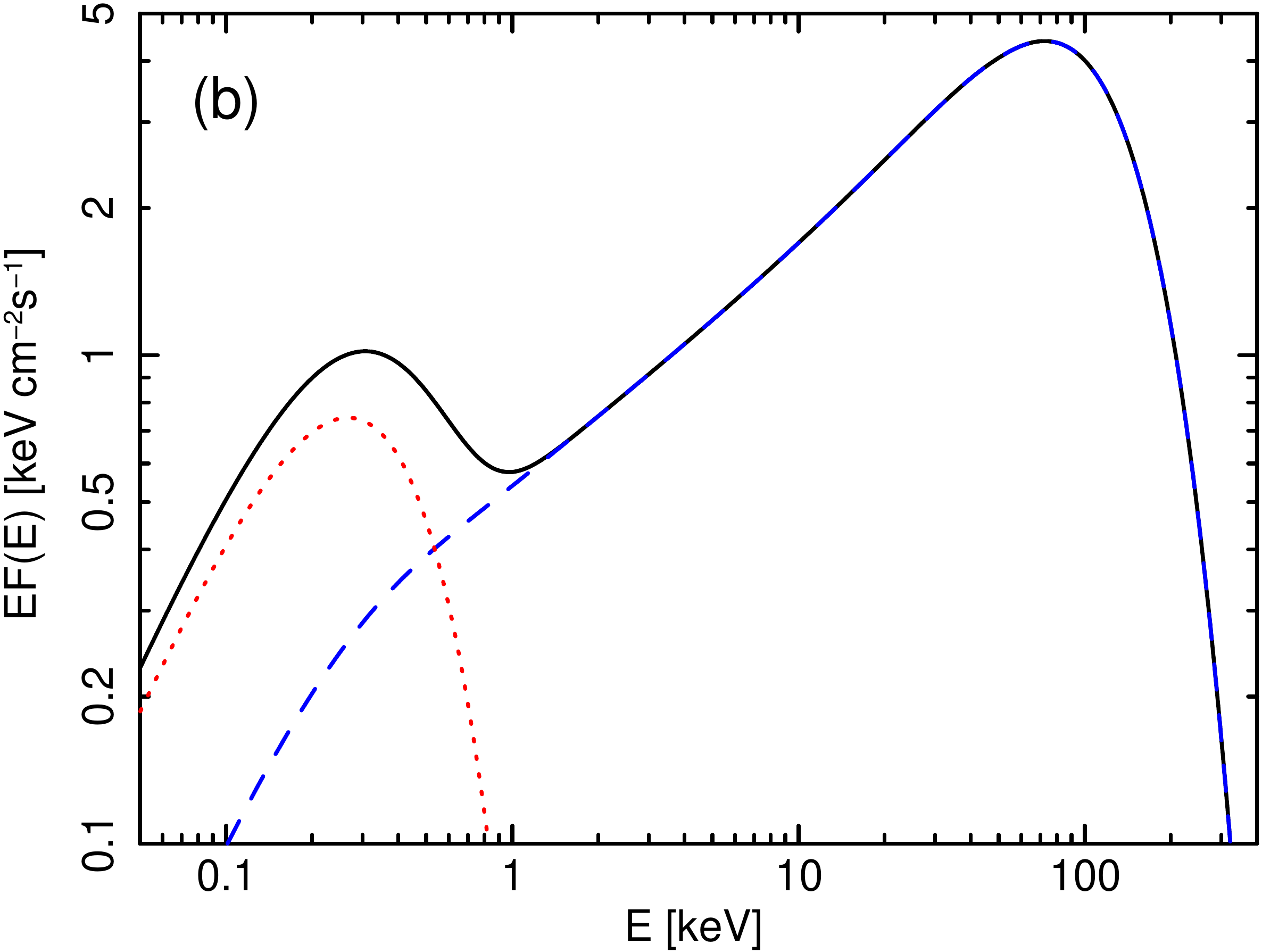}
  \includegraphics[height=6.cm,angle=-90]{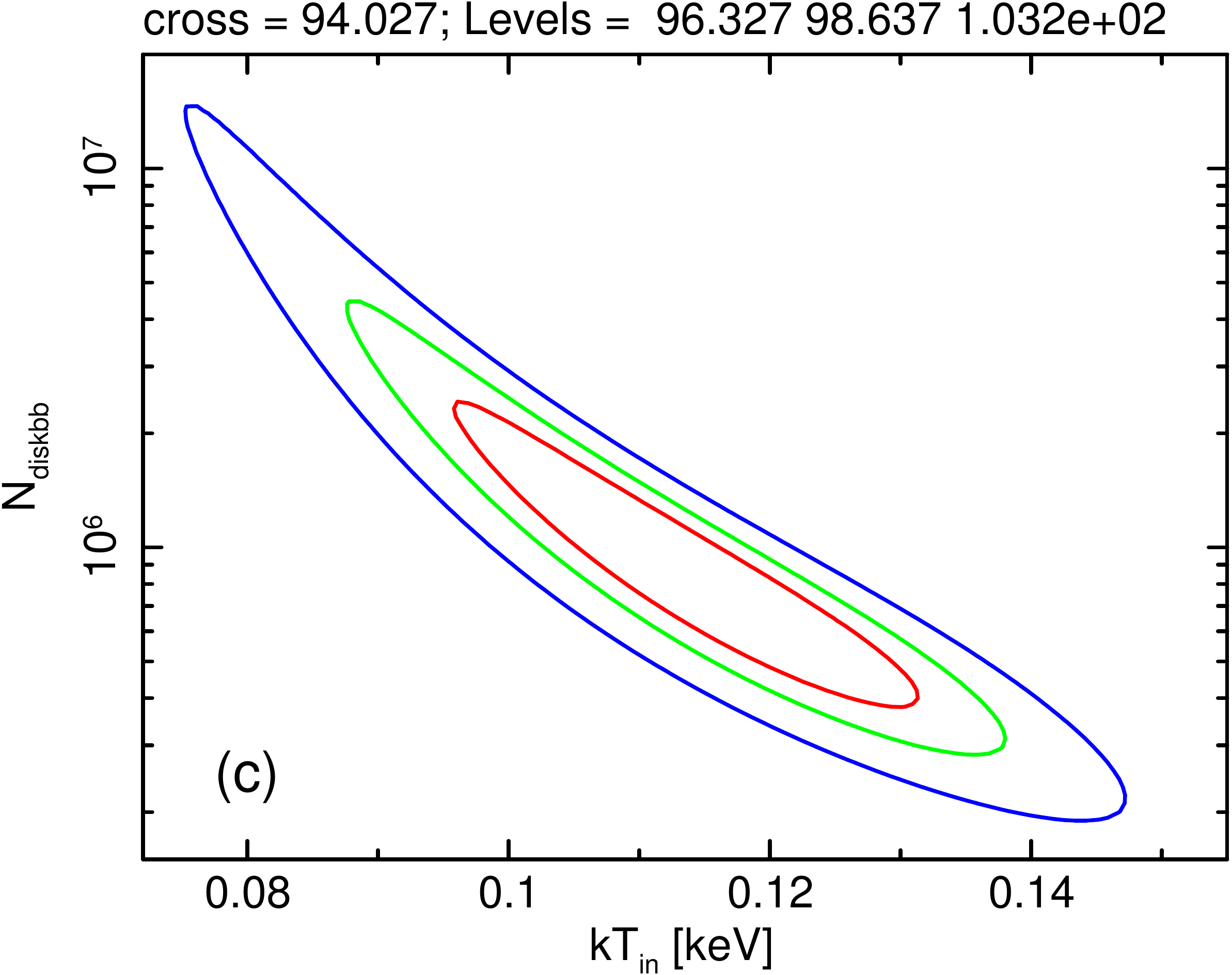}
  \caption{(a) The XRT spectrum fitted in the 0.55--6\,keV with thermal Comptonization of a fraction of disk blackbody photons (and absorbed in the ISM). The upper and lower panels show the unfolded spectrum and the data/model ratio. In addition, the data require two Gaussian lines to account for instrumental artifacts, shown by the doted curves. (b) The corresponding unabsorbed model. The red dots and blue dashes show the unscattered part of the disk blackbody and the Comptonization, respectively, and the solid black curve shows the sum. (c) The 1, 2 and 3$\sigma$ confidence contours for $N_{\tt diskbb}$ and $kT_{\rm in}$.
}\label{XRT}
\end{figure}

\setlength{\tabcolsep}{3pt}
\begin{table}
\centering            
\caption{The results of the spectral fit of the XRT spectrum in the 0.55--6 keV range (107 channels) assuming $kT_{\rm e}=30$\,keV with {\tt{jscrab*tbabs*thcomp(diskbb)}}.}
\label{XRT_fit}      
\vskip -0.4cm                               
\begin{tabular}{ccccccc}
\hline 
$N_{\rm H}$ & $\Gamma$ & $f_{\rm sc}$ & $kT_{\rm in}$  
 & $N_{\tt diskbb}$& $\chi_\nu^2$ \\
$10^{21}{\rm cm}^{-2}$ &  &  &keV & $10^6$ &   \\
\hline
$6.5_{-0.3}^{+0.4}$ & $1.52^{+0.03}_{-0.02}$ & $0.24_{-0.10}^{+0.12}$ & $0.11_{-0.02}^{+0.02}$ & $0.9_{-0.5}^{+1.9}$ & 94/96\\
\hline                    
\end{tabular} 
\end{table}

We first consider the XRT spectrum. We fit it in the 0.55--6 keV range with the thermal Comptonization model of \citet{Z20_thcomp}, {\tt thcomp}. This is a convolution model, allowing for Comptonization of a fraction of photons from any form of the seed spectrum. We assume the seed photons have a disk blackbody spectrum ({\tt diskbb}; \citealt{Mitsuda84}). (Adding {\tt diskbb} to an absorbed power law model has the probability of 10$^{-4}$ of being by chance.) As stated above, we account for the spectral differences between the instruments using the {\tt jscrab} model, where we fix $\Delta \Gamma$ at 0.077 and the normalization at 0.83 as found in our overall best-fit model below for the joint XRT/PCA/HEXTE data (which assume a power-law irradiation profile), see Section \ref{joint}.

We obtain a very good fit, whose parameters are given in Table \ref{XRT_fit}. It shows the XRT spectrum can be self-consistently accounted for by Compton scattering of a relatively small fraction of the disk blackbody photons. Figures \ref{XRT}a, b show the unfolded spectrum with the absorbed model and the data/model ratio, and the unabsorbed model components, respectively. Since the two parameters of the disk blackbody are anticorrelated, we show their confidence contours on Figure \ref{XRT}c. We see that the power-law-like component dominates down to $\approx$1\,keV, while the thermal disk has a rather low temperature. The fitted value of $N_{\rm H}$ is consistent with that of the Galactic column density. We have also tested the effect of switching to the abundances of \citet{WAMC00}. Then, $N_{\rm H}\approx 9.5^{+0.5}_{-0.4}\times 10^{21}$\,cm$^{-2}$, $f_{\rm sc}\approx 0.11^{+0.06}_{-0.05}$, $kT_{\rm in}\approx 0.10\pm 0.01$\,keV, and the $N_{\tt diskbb}$ is significantly larger, $\approx 3.0^{+5.9}_{-1.8}\times 10^6$.
 
We note our value of $kT_{\rm in}$ is different from that of \munoz, $kT_{\rm in}\approx 0.313\pm 0.007$\,keV. They used the sum of disk blackbody and a power law, and we thus fitted the XRT data with that model. We obtain $kT_{\rm in}\approx 0.10^{+0.02}_{-0.02}$\,keV, very similar to that of the Comptonization model described above. They also used the average of the first three XRT observations only. We repeated this procedure, and also found $kT_{\rm in}\lesssim 0.12$\,keV. This also confirms that the obtained low value of $kT_{\rm in}$ is not an artifact of the summing of the individual XRT spectra. We argue that the value given in \munoz might to be a typo considering that their {\tt diskbb} normalization ($\approx 1.0\times 10^6$) at $kT_{\rm in}\approx 0.3$\,keV yields a disk blackbody flux ($\propto T_{\rm in}^4$) $\sim$100 times higher than that of the data (which is $\sim\! 10^{-8}$\,erg\,cm$^{-2}$s$^{-1}$). This discrepancy disappears for $kT_{\rm in}\approx 0.1$\,keV.

As we discuss in Section \ref{constraints}, the fitted disk blackbody normalization implies the disk inner radius of $\sim\! 10^2 R_{\rm g}$. If the abundances of \citet{WAMC00} are used, even larger values are implied. Either is completely incompatible with the inner radius of $\lesssim 2 R_{\rm g}$ found in \garcia. Furthermore, the low inner disk temperature we find implies a large truncation radius as well (Section \ref{constraints}). Below, we include the \xte data in our study, considering possible ways to reconcile them with the findings based on the XRT data.

\subsection{Analysis of joint spectra}
\label{joint}

\begin{table*}
\centering            
\caption{The list of main broad-band models in Section \ref{joint}. All of the models begin with {\tt jscrab*tbabs}, and include instrumental lines for the PCA and XRT data (not shown).
}
\label{models}      
\vskip -0.4cm                               
\begin{tabular}{|l|l|l|}
\hline 
\# & {\sc xspec} structure & Description\\
\hline 
1 & ({\tt relxilllpCp+simplcut(relxilllpCp)} & PCA/HEXTE, following 3.A of \garcia, reflection $\ll$ lamppost\\
& {\tt +xillverCp})&\\
2 & ({\tt reflkerr\_lp+hreflect}) & PCA/HEXTE, bottom lamp, reflection = lamppost \\
3 & ({\tt diskbb+reflkerr\_lp+hreflect}) & XRT/PCA/HEXTE, bottom lamp, reflection = lamppost, additional disk blackbody\\
4 & ({\tt diskbb+reflkerr+hreflect}) & XRT/PCA/HEXTE, coronal model with a power-law profile + disk blackbody\\
5 & ({\tt reflkerr\_lpbb+hreflect}) & XRT/PCA/HEXTE, bottom lamp, reflection = lamppost, quasi-thermal disk emission\\
\hline                   
\end{tabular} 
\tablecomments{All of the lamppost models with satisfactory fits violate e$^\pm$ pair equilibrium, have most of the photons emitted by the lamppost trapped by the BH, and do not fully account for the re-emission of the irradiating flux. The coronal model has an extremely steep irradiation profile. Thus, neither model can be considered physical.} 
\end{table*}

We repeat the analysis of \citetalias{Garcia18} for their model 3.A of the joint PCA/HEXTE spectrum, which is one of their two best models (the other assumes the disk inner radius exactly at the ISCO, $R_{\rm{in}}= R_{\rm{ISCO}}$). It assumes a lamppost above a rotating BH with the dimensionless spin of $a_*=0.998$ (for which $R_{\rm{ISCO}}\approx 1.237 R_{\rm{g}}$ and $R_{\rm hor}\approx 1.063 R_{\rm{g}}$, where $R_{\rm g}\equiv GM/c^2$). The incident spectrum is from thermal Comptonization, using the model {\tt nthcomp}. Their model also includes thermal Comptonization of a fraction of the reflected emission by the electrons in the lamppost, using the {\tt simplcut} model of \citet{Steiner17} and a weak contribution of static reflection from remote parts of the disk (using {\tt xillverCp}), as given in the {\sc xspec} notation in Table \ref{models} as Model 1. We used two {\tt relxilllpCp} components to describe the primary and reflected emission separately, respectively, see the first and second components in Table \ref{models}. We note here that {\tt relxilllpCp} gives the electron temperature, $kT_{\rm e}$, in the observer's frame\footnote{\url{http://www.sternwarte.uni-erlangen.de/~dauser/research/relxill/}}, i.e., redshifted with respect to that in the lamppost by $1+z$, which is given by
\begin{equation}
1+z=\sqrt{\frac{(H/R_{\rm g})^2+a_*^2}{(H/R_{\rm g})^2+a_*^2-2 H/R_{\rm g}}}.
\label{redshift}
\end{equation}
We thus set the temperature of the scattering electrons in {\tt simplcut} to $(1+z)k T_{\rm e}$, while the observer's frame $kT_{\rm e}$ was used in \garcia (J. Garc{\'{\i}}a 2020, private communication). \garcia assumed the abundances of \citet{WAMC00}, which we also assume in this model. With the current version (1.3.5) of the {\tt relxill} family of codes, we find an excellent fit with this model, even better than that reported in \citetalias{Garcia18}, with $\chi_\nu^2\approx 101/90$, compared to their 117/89. We obtain lower relativistic blurring than \citetalias{Garcia18}, with our best-fit lamppost height at $H\approx 2.88_{-0.53}^{+0.23} R_{\rm hor}$ (vs.\ $1.17 R_{\rm hor}\approx 1.24 R_{\rm g}$ in \garcia), $R_{\rm in}\approx 2.13^{+0.16}_{-2.13} R_{\rm ISCO}$ (vs.\ $1.8 R_{\rm ISCO}$ in \garcia), $i\approx 25\pm 2\degr$, and $N_{\rm H}\approx 1.06^{+0.04}_{-0.05}\times 10^{22}$\,cm$^{-2}$. However, our new fit has some self-consistency problems. First, $0.40^{+0.06}_{-0.20}$ of the disk-reflected emission directed toward the observer undergo subsequent scattering in the primary Comptonizing source, which is approximated as a point source. As shown by \citet{DD16}, much fewer reflected photons would go back to the lamppost, especially given the fitted low disk inclination. Second, the fitted reflection fraction (defined as in \citealt{Dauser16}) is $R_{\rm f}\approx 0.50^{+0.06}_{-0.04}$, while the value predicted by the lamppost model is $\approx$2.08 (as obtained by setting {\tt fixReflFrac} = 2 in the model), i.e., about four times more. 

Thus, we modify Model 1 by setting the reflection fraction equal the lamppost value. Since the PCA data are for $\geq$3\,keV only, the fitted Galactic absorption column may be not fully correct as well. Therefore, we also impose $N_{\rm H}=6.5\pm 0.5\times 10^{21}$\,cm$^{-2}$ and switch back to the abundances of \citet{AG89}, based on the fitting the XRT data. However, we have found we are then unable to obtain a good fit, with the best found $\chi_\nu^2$ of only $\approx 174/91$. 

Then, we switched to the {\tt reflkerr\_lp} model of \citet{Niedzwiecki19}, shown as Model 2 in Table \ref{models}, where the static reflection component {\tt hreflect} (\citealt{Niedzwiecki19}; combining {\tt{xillver}} and {\tt ireflect}, \citealt{MZ95}) accounts for reflection from remote parts of the disk, assumed to be close to neutral. The {\tt reflkerr\_lp} has some advantages over {\tt relxilllpCp}: it includes the emission of the bottom lamppost, it models the incident Comptonization photon using the {\tt compps} model of \citet{PS96} (which is significantly more accurate than the {\tt nthcomp} model used in {\tt relxilllpCp}), and it uses a correct relativistic treatment of Compton reflection at energies $\gtrsim$10\,keV, while {\tt relxilllpCp} uses a non-relativistic treatment \citep{GK10}. At low inclination angles, the gravitationally focused emission from the bottom lamppost enhances the direct emission \citep{NZ18}, which then reduces the fractional reflection, which is equivalent to artificially reducing the reflection fraction from its physical value when only the top lamp emission is included (as in {\tt relxilllpCp}). This effect apparently prevented finding a good fit with the above {\tt relxilllpCp} model at the physical reflection normalization. Furthermore, we find that the fit with {\tt reflkerr\_lp} no longer requires any scattering of the reflected emission. This has been tested using the {\tt thcomp} model (instead of {\tt simplcut}), which is compatible with {\tt compps} (see \citealt{Z20_thcomp}) used as the incident spectrum in {\tt reflkerr\_lp}. The incident Comptonization spectrum includes only the scattered photons, since it is emitted by electron scattering in the lamppost. Since the {\tt reflkerr\_lp} model gives both $kT_{\rm e}$ and $kT_{\rm seed}$ in the lamppost frame whereas {\tt hreflect} has parameters almost equal to those in the observer's frame, we account for the gravitational redshift of the direct lamppost emission by setting the electron and seed temperatures in {\tt hreflect} equal to $kT_{\rm e}/(1+z)$, $kT_{\rm seed}/(1+z)$. We obtain an excellent fit with $\chi_\nu^2\approx 102/91$, and $H\approx 1.71_{-0.11}^{+0.10} R_{\rm g}\approx 1.61 R_{\rm hor}$, $R_{\rm in}\approx 3.18^{+0.25}_{-0.30} R_{\rm ISCO}\approx 3.93 R_{\rm g}$, $N_{\rm H}\approx 6.6^{+0.5}_{-0.7}\times 10^{21}$\,cm$^{-2}$ (compatible with the constraint from the XRT spectrum). In this model, the bottom lamp is fully visible, while it is likely to be partially obscured by the fast flow below $R_{\rm in}$. We have tested it, and found that the 90\% confidence range of the fraction of the bottom-lamp emission being visible is $\delta\geq 0.58$. 

Similarly to the previous models for the PCA/HEXTE spectrum, the reflecting medium is found to be strongly ionized, with $\log_{10}\xi\approx 3.49^{+0.19}_{-0.10}$, where the ionization parameter is defined as
\begin{equation}
\xi\equiv 4{\pi}F_{\rm{irr}}/n,
\label{xi}
\end{equation}
where $F_{\rm{irr}}$ is the irradiating flux measured at the source in the 0.1--1000\,keV photon energy range (J. Garc{\'\i}a 2020, private communication) and $n$ is the electron density. For our fitted spectra, this energy range contains most of the bolometric flux, and we hereafter neglect the small inaccuracy resulting from not accounting for the flux beyond this range.

However, we find a number of issues with this model. The gravitational focusing of the bottom lamppost implies a strongly fine-tuned (and very low) inclination, $i\approx 9.7^{+0.4}_{-0.4}\degr$. Then, 72\% of the primary photons are captured by the BH. Thus, this model has the radiative efficiency more than three times lower than that of models with negligible photon capture by the BH. This needs to be compensated by increasing the accretion rate, $\dot M$, by a factor of $\gtrsim$3, which, as can be inferred from fig.\ 1 of \garcia, would make the $\dot M$ in the LHS higher than that in the following soft state (which effect was pointed out by \citealt{NZS16}). In the soft state, most of the X-ray flux is from the intrinsic emission from the disk extending to the ISCO, which is much less affected by the photon trapping than the emission of the lamppost. The electron temperature is $kT_{\rm e}\approx 174^{+17}_{-17}$\,keV, for which the Comptonization spectrum peaks around the threshold for e$^\pm$ pair production, 511\,keV. The compactness parameter in the lamppost frame for the the fitted height and including all photons is $\ell\approx 2.1\times 10^4$, and that for only photons above the threshold for pair production is $\ell(>511\,{\rm keV})\approx 2.7\times 10^3$. Here, the compactness parameter is defined by 
\begin{equation}
\ell \equiv \frac{L_{\rm intr}\sigma_{\rm T}}{D m_{\rm e}c^2},\quad D=\frac{H-R_{\rm hor}}{2},
\label{ell}
\end{equation}
where $L_{\rm intr}$, $\sigma_{\rm T}$ and $m_{\rm e}c^2$ are the lamppost luminosity in the its frame, the Thomson cross section and the electron rest energy, respectively, while the lamppost size, $D$, has been estimated based on its distance to the horizon. As shown, e.g., in fig.\ 1 of \citet{Fabian15} (see also \citealt{Zdziarski85, Stern95}), we expect runaway e$^\pm$ pair production at the above $kT_{\rm e}$ for $\ell\gtrsim 10$, i.e., our fit strongly violates the pair equilibrium. Finally, our model has a rather high Fe abundance, $Z_{\rm Fe}\approx 5.3_{-0.8}^{+1.9}$, also unlikely to be real. We note that similar problems occur for the original model of \garcia as well as for our other lamppost models in this section. 

Still, given that the model provides a good phenomenological description of the broad-band model, we apply it to the joint XRT, PCA and HEXTE data, see Model 3 in Table \ref{models}. Since these data extend down to 0.55\,keV, disk blackbody photons and the seed photons for Comptonization become important. We account for the former using the {\tt diskbb} model, in which the innermost disk temperature, $kT_{\rm in}$, is given in the observer's frame. Disk photons undergoing Comptonization in the lamppost are blueshifted, for which we take account by assuming the seed photon temperature of $T_{\rm seed}=(1+z)T_{\rm in}$. Most of those photons are from close to the disk inner edge, and we therefore assume the seed photons have a blackbody distribution. As for the PCA/HEXTE model, we find a good fit with similar parameters (and thus the same problems as discussed above) at $\chi^2\approx 204/192$, and $R_{\rm in}\approx 3.4\pm 0.3 R_{\rm ISCO}$. The parameters of the disk blackbody are similar to those of the model for the XRT alone, $k T_{\rm in}\approx 0.11^{+0.02}_{-0.01}$\,keV, $N_{\tt diskbb}\approx 5^{+8}_{-1}\times 10^5$. We note that this component is introduced purely phenomenologically, and we do not yet address the issue whether the inner disk irradiated by a luminous lamppost can keep such a low inner temperature (see Section \ref{quasi} below). The XRT and PCA data appear to be in a relatively good agreement in the overlapping range of 3--6\,keV (after correcting for the slope and normalization differences with {\tt jscrab}).

\begin{figure}
  \centering
  \includegraphics[height=7.5cm,angle=-90]{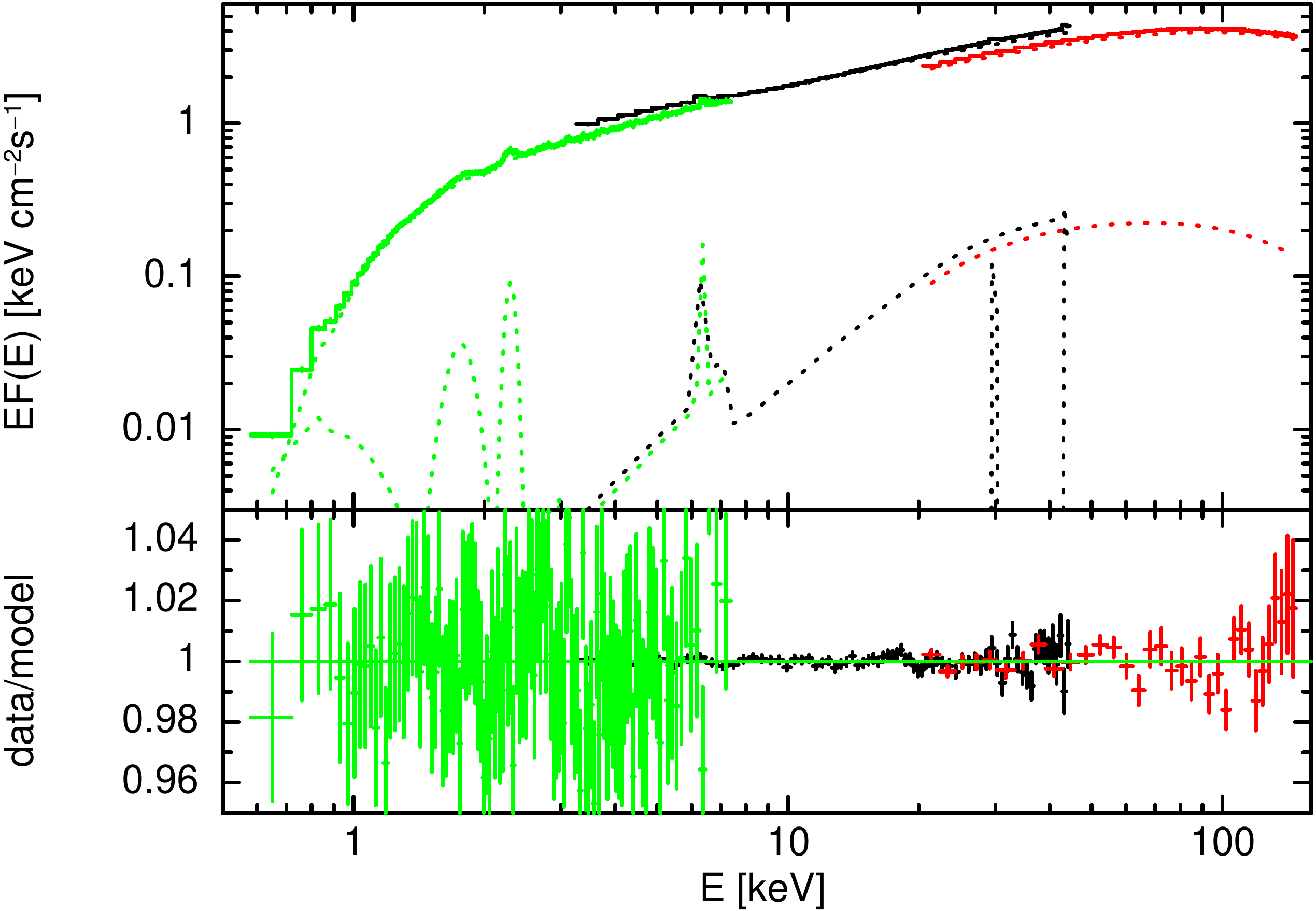}
  \caption{The XRT, PCA and HEXTE spectra as fitted in the 0.55--140\,keV with a disk blackbody and relativistic coronal reflection, Model 4 in Table \ref{models}, shown as the unfolded spectra and the data/model ratios. The green, black and red symbols correspond to the XRT, PCA and HEXTE data, respectively. The dotted curves in the upper panel show the model components: the disk blackbody, the sum of the incident and relativistically reflected spectra, and the remote reflection. In addition, two lines account for each of the XRT and PCA instrumental residuals at $\approx$1.7, 2.3\,keV, and $\approx$30, 44\,keV, respectively.
}\label{coronal}
\end{figure}

We have also fitted the joint XRT, PCA, HEXTE spectra by a coronal model with a phenomenological power-law radial irradiation profile, see Model 4 in Table \ref{models} and Figure \ref{coronal}. However, this model, while yielding the lowest $\chi^2_\nu$ among the considered models to the XRT/PCA/HEXTE data, $\approx 198/192$, requires an extremely steep profile, $\propto R^{-q}$, $q\approx 15.0^{+1.9}_{-1.4}$, as well as the disk extending to the ISCO, $R_{\rm in}\approx 1.00^{+0.09} R_{\rm ISCO}$, and a very high inclination, $i=78^{+1}_{-7}\degr$. Such a steep irradiation profile is unphysical \citep{NZS16}, and the high inclination unlikely. The reflection component is relatively weak, ${\cal R}\approx 0.26^{+0.01}_{-0.02}$, $0.18^{+0.02}_{-0.04}$ for the relativistic and remote reflection, respectively. Hereafter, the reflection fraction, $\mathcal{R}$, is defined as the ratio of the flux locally emitted by the primary source toward the reflector to that away from it, e.g., $\mathcal{R}=1$ for an isotropic source above a slab. The scattering plasma has the parameters of $\Gamma\approx 1.51$, $kT_{\rm e}\approx 66^{+3}_{-3}$\,keV, $kT_{\rm in}\approx 0.12^{+0.01}_{-0.02}$\,keV. The bolometric unabsorbed flux of this model is $\approx 3\times 10^{-8}$\,erg\,cm$^{-2}$\,s$^{-1}$, and the flux in the Comptonization component is $F_{\rm C}\approx 2.0\times 10^{-8}$\,erg\,cm$^{-2}$\,s$^{-1}$.

\subsection{Quasi-thermal disk emission}
\label{quasi}

The above fits neglect quasi-thermal radiation of the disk due to both re-emission of the power absorbed within the disk and the emission from the likely viscous dissipation (with the flux $F_{\rm diss}$). The latter effect is neglected in the {\tt reflkerr\_lp} (and {\tt relxilllpCp}) model. The former is also effectively neglected because the spectrum below 10 keV is modelled using {\tt xillver} (\citealt{GK10, Garcia13}; \garcia), which assumes the reflector density of $n=10^{15}$\,cm$^{-3}$. At the ionization parameter fitted to the XRT/PCA/HEXTE data of $\log_{10}\xi\approx 3.41^{+0.09}_{-0.05}$, the flux irradiating the disk is (Equation \ref{xi}) $F_{\rm irr}= \xi n/(4\pi) \approx 2.1^{+0.4}_{-0.3}\times 10^{17}$\,erg\,cm$^{-2}$\,s$^{-1}$. This is several orders of magnitude below the flux irradiating parts of the disk close to the ISCO estimated (following the method of \citealt{ZDM20}) using the possible ranges of the distance and the BH mass of \source. As required by the Stefan-Boltzmann law, the re-emission of the absorbed power occurs at energies higher than those of a blackbody at the effective temperature of the irradiated medium, $T_{\rm{eff}}$, 
\begin{equation}
\sigma T_{\rm{eff}}^4=(1-a)F_{\rm{irr}}+(1 - f_{\rm c})F_{\rm{diss}},
\label{teff}
\end{equation}
where $a$ is the albedo for backscattering, $f_{\rm c}\leq 1$ is the fraction of the internal dissipation power transferred at $R>R_{\rm in}$ to a corona or a lamppost \citep{SZ94}, and $\sigma$ is the Stefan-Boltzmann constant. Neglecting $F_{\rm{diss}}$ and scaling the albedo to $a=0.5$ typical for reflection from ionized media, we have $k T_{\rm eff}\approx 18\pm 1[(1-a)/0.5]^{1/4}$\,eV using the above $F_{\rm irr}$. Consequently, a quasi-thermal re-emission at this $T_{\rm eff}$ can only marginally affect the fitted energy range of 0.55--140\,keV. 

The above effects are approximately taken into account in the high-density version of the static {\tt reflionx} radiative transfer code \citep{Ross99, Ross07, Tomsick18}, which, when convolved with a code from the {\tt relconv} family \citep{Dauser10, Dauser13}, can describe relativistic reflection/reprocessing spectra in either coronal or lamppost geometry. However, that version of {\tt reflionx} is not publicly available. Therefore, in order to account for these effects in the disk-lamppost geometry, we have developed a new model, {\tt reflkerr\_lpbb}, described in detail in Appendix \ref{thermal}. The irradiating flux at each disk radius is calculated exactly in the lamppost geometry, with the free parameters being $M$ and $d$. The intrinsic, dissipative, flux is calculated using the GR disk model of \citet{NT73} scaled by $(1-f_{\rm c})$ (to account for the reduction of the disk emission due to the transfer of energy away from the disk). That model assumes a zero-stress inner boundary condition at the ISCO. The disk accretion rate is calculated from the lamppost luminosity and the accretion efficiency. Then, the disk emission is approximated as a diluted blackbody integrated over the disk surface (and taking into account GR effects) at the color temperatures of $T_{\rm col}\equiv \kappa T_{\rm eff}$, where $\kappa\approx 1.3$--1.7 \citep{Davis05} is the color correction factor. While using a diluted blackbody is only a rough approximation to the actual reflection spectra, it does reproduce their overall shape, see, e.g., fig.\ 1 in \citet{ZDM20}.

Model 5 in Table \ref{models} includes {\tt reflkerr\_lpbb}. We assume the lowest plausible values of the distance of $d=4$\,kpc and $\kappa=1.3$, and the highest plausible BH mass in a low-mass X-ray binary, $M=15\msun$. We first allow free $f_{\rm c}$ and $a$ (which parameter here corresponds to the fraction of the incident flux re-emitted below 0.1\,keV, see Appendix \ref{thermal}). In this model, we have no {\tt diskbb} component, since its role is now taken over by the diluted blackbody disk emission. In the assumed geometry, a fraction of this emission is then Comptonized in the lamppost, resulting in the incident spectrum. However, taking it into account fully self-consistently would require a convolution version of Comptonization in {\tt reflkerr\_lpbb}, which is at present not available. Therefore, we allow the seed photon temperature for Comptonization to be free. We have obtained a relatively good fit, with $\chi^2_\nu\approx 205/192$ and the remaining parameters almost identical to those in the lamppost model without the blackbody component, but the obtained albedo parameter is $a\approx 1.00_{-0.10}$. However, the actual albedo parameter of the fitted spectrum can be calculated by averaging the reflected spectra over $\cos i$, which yields $a\approx 0.56$, almost independent of the radius. Then, the fit of the model with the fixed $a=0.56$ becomes much worse, with $\chi^2_\nu\approx 237/193$. This is due to pronounced residuals in the XRT energy range related to the predicted (but not present) quasi-thermal component. Its incident spectrum has even higher electron temperature, $kT_{\rm e}\approx 300$\,keV and the compactness parameter is $\ell\approx 4.9\times 10^4$, which strongly violate the pair equilibrium. In addition, 7 times more photons falls into the BH than escapes, which would then imply the hard-state luminosity to be larger than the soft-state one.

Summarizing, we have been able to find good phenomenological models for the broad-band spectrum, either including a {\tt diskbb} component or without it but including some intrinsic disk emission. However, those models are unambiguously unphysical. They strongly violate the e$^\pm$ pair equilibrium, have an implausibly low radiative efficiency due to photon trapping by the BH, and either include an ad hoc disk blackbody component at a very low $kT_{\rm in}$ or, in the model with intrinsic disk emission and re-radiation, require the albedo parameter to be close to unity, while the actual value is close to a half. 

In addition, all of the models show the reflection component to be weak, weaker than $\approx$1/4 of that corresponding to a lamppost illuminating a surrounding disk. In the original model of \garcia, this problem was solved by assuming that 83\% of the reflection is then upscattered by hot electrons in the lamppost, which is clearly geometrically impossible \citep{DD16}. In our variant of the lamppost model, this is solved by invoking the gravitationally focused emission of the bottom lamp, which then works only if the inclination is within a narrow range around $\approx\!10\degr$, requiring fine tuning. The observed reflection can be weak in the coronal geometry, as indeed found in our coronal model, but that broad-band model requires an extreme irradiation profile, $q\approx 15.0^{+1.9}_{-1.4}$, which cannot be physical. A likely solution to the above problems is an inhomogeneity of the primary X-ray source, as we discuss below.

\subsection{Inhomogeneity of the X-ray source}
\label{inhomogeneity}

\begin{figure}
  \centering
  \includegraphics[width=7cm]{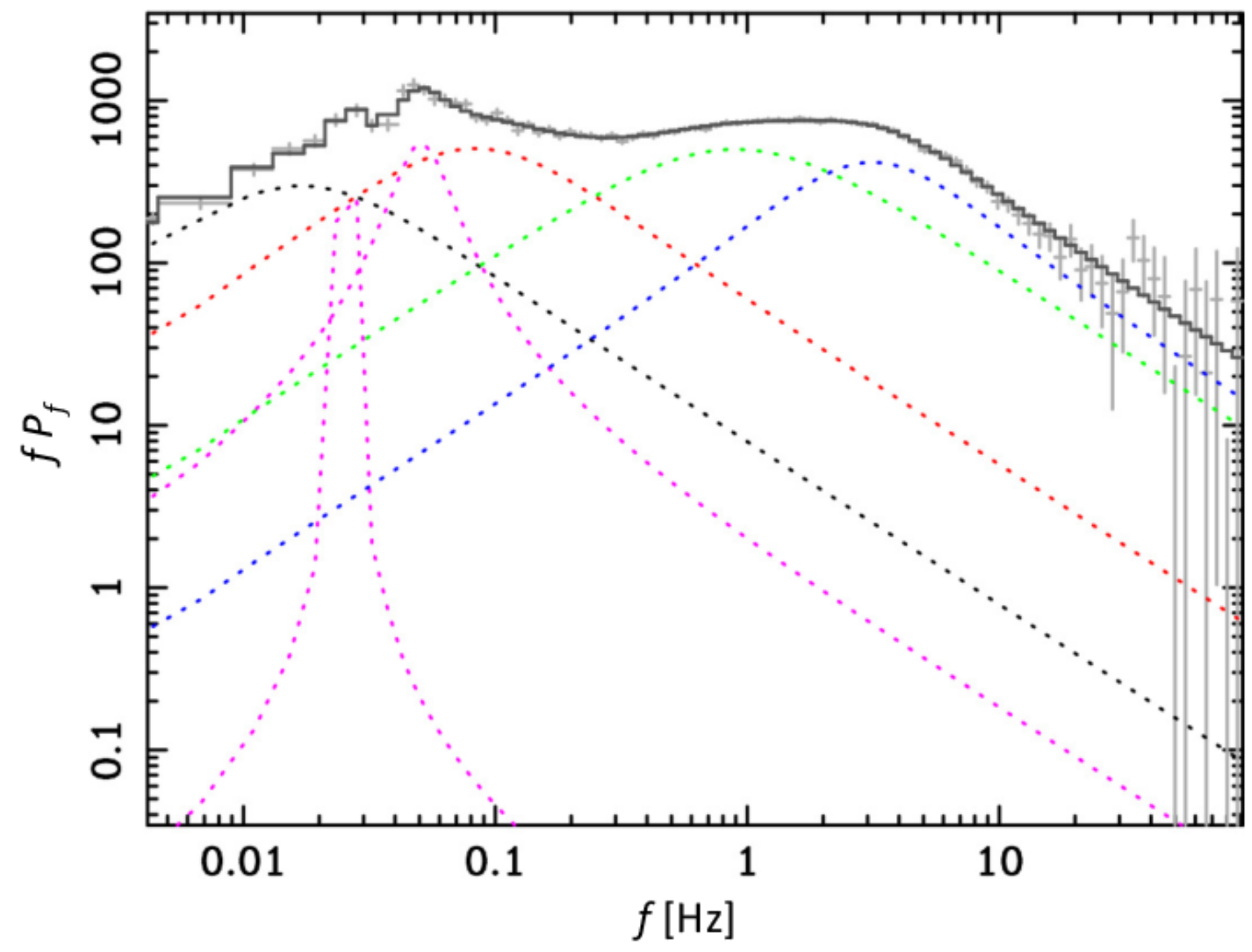}
  \caption{The power spectra of the PCA observations during MJD 55130--55132 fitted with four broad and two narrow Lorentzian components.
}\label{PSD_model}
\end{figure}

The applicability of models with a single-component Comptonization reflecting from a surrounding disk to \source is also at odds with results from its X-ray variability studies, which support an inhomogeneous structure. \munoz studied time variability of the first three days of the same PCA data we consider here. They found a complex power spectrum, whose fitting required several separate Lorentzian components. We have obtained the individual power spectra for the entire PCA observation considered here. We have found they can be grouped into three epochs, each corresponding to periods during which the source was stationary, namely MJD 55130--55132, 55133--55148, 55149--55155. Figure \ref{PSD_model} shows the first one. The spectrum has indeed a complex shape, which fitting requires six Lorentzian components. We consider it highly unlikely that this complexity is generated in a lamppost irradiating a surrounding disk. \munoz also found pronounced time lags of hard X-rays vs.\ soft ones, dependent on the Fourier frequency as $f^{0.7}$, which also argue for the primary source being extended in the equatorial plane. 

\munoz pointed out the remarkable similarity of the LHS of this source to that of Cyg X-1. For that object, we know well that the Fourier-resolved spectra do strongly depend on the frequency and the primary source appears to be composed of three main Comptonization components \citep{Axelsson18,Mahmoud18b}. In particular, \citet{Mahmoud18b} show the dominance of the hardest spectral component at $\gtrsim$10\,keV.

Given the likely inhomogeneity of the source, we have tried to fit different parts of the broad-band spectrum with simple models. In this way, we have found we could fit the part of the PCA spectrum at $\lesssim$10\,keV with very simple models. This range is crucial for precise determination of the relativistic broadening of the reflection, as it includes the intrinsically narrow Fe K lines as well as their wings, and the Fe K edge. The range above 10 keV includes the reflection hump, which is very broad, and thus weakly affected by relativistic effects, as well as much more difficult to distinguish from the primary continuum. We also stress that the 3.3--10.1\,keV energy range still contains most of the PCA counts, with $6.4\times 10^7$ of out the total of $1.0\times 10^8$ in 17 channels. 

If this source is indeed similar to Cyg X-1, this part of the spectrum is likely to be dominated by the outer Comptonizing regions, producing a softer spectrum. Indeed, we find we can achieve a very good fit to those data with a model with a broken-power law incident spectrum and reflection from a static neutral medium at the solar abundances using an absorbed {\tt bexrav} model \citep{MZ95} and a narrow Gaussian line at 6.40 keV (as fitted if allowed free), obtaining $\chi_\nu^2 \approx 7/10$. The power law breaks from $\Gamma\approx 1.74_{-0.09}^{+5.75}$ below $3.8_{-0.5}^{+0.3}$\,keV to $\Gamma\approx 1.56_{-0.01}^{+0.01}$ above it. The reflection is weak, $\mathcal{R}\approx 0.22^{+0.07}_{-0.07}$. In this model, the disk inclination has been fixed at $30\degr$. However, we can achieve equally good fits at any inclination ($\Delta \chi^2\leq 0.35$ for $i\leq 87\degr$). 

\begin{figure}
  \centering
  \includegraphics[height=6.5cm,angle=-90]{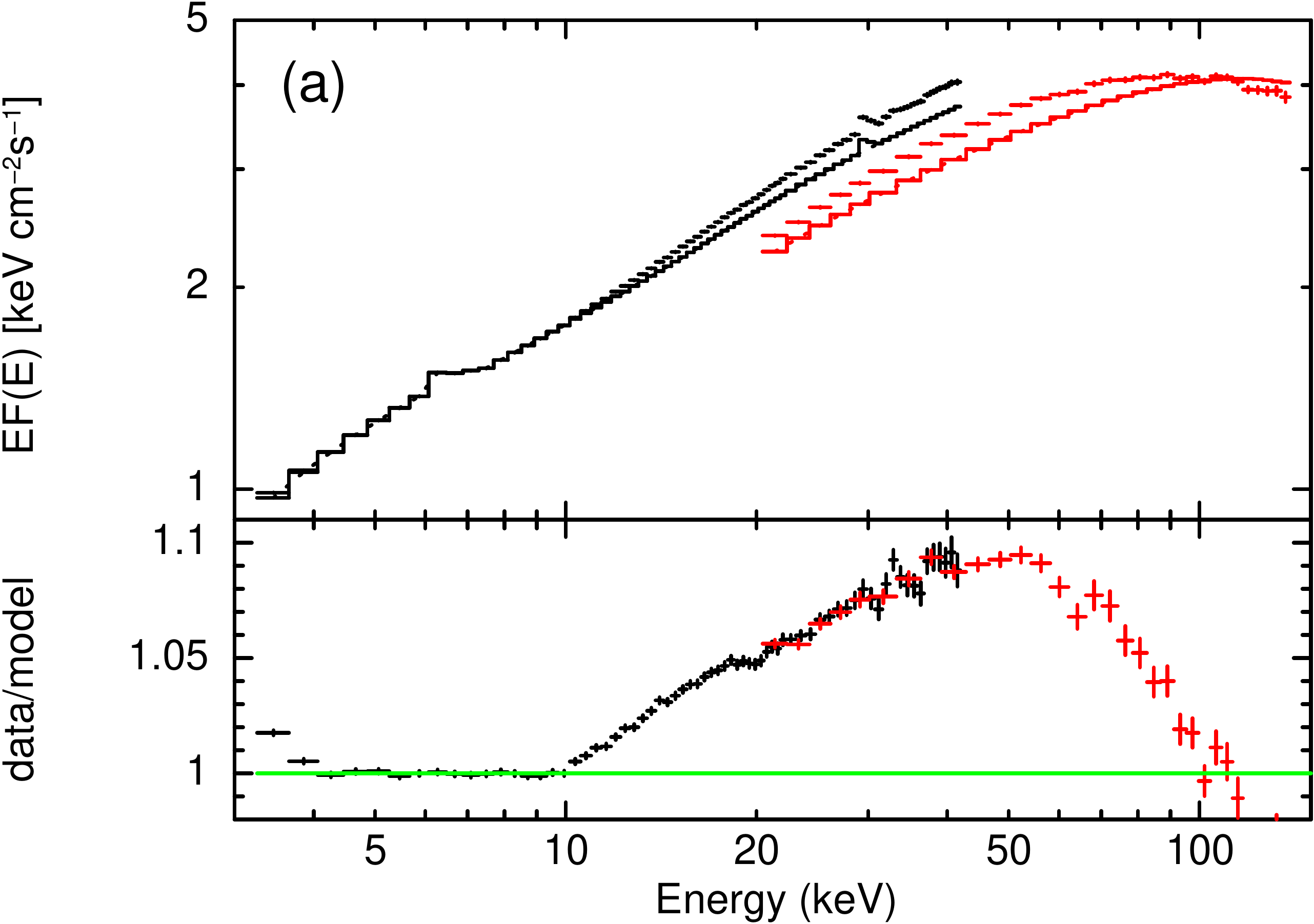}
  \includegraphics[height=6.cm,angle=-90]{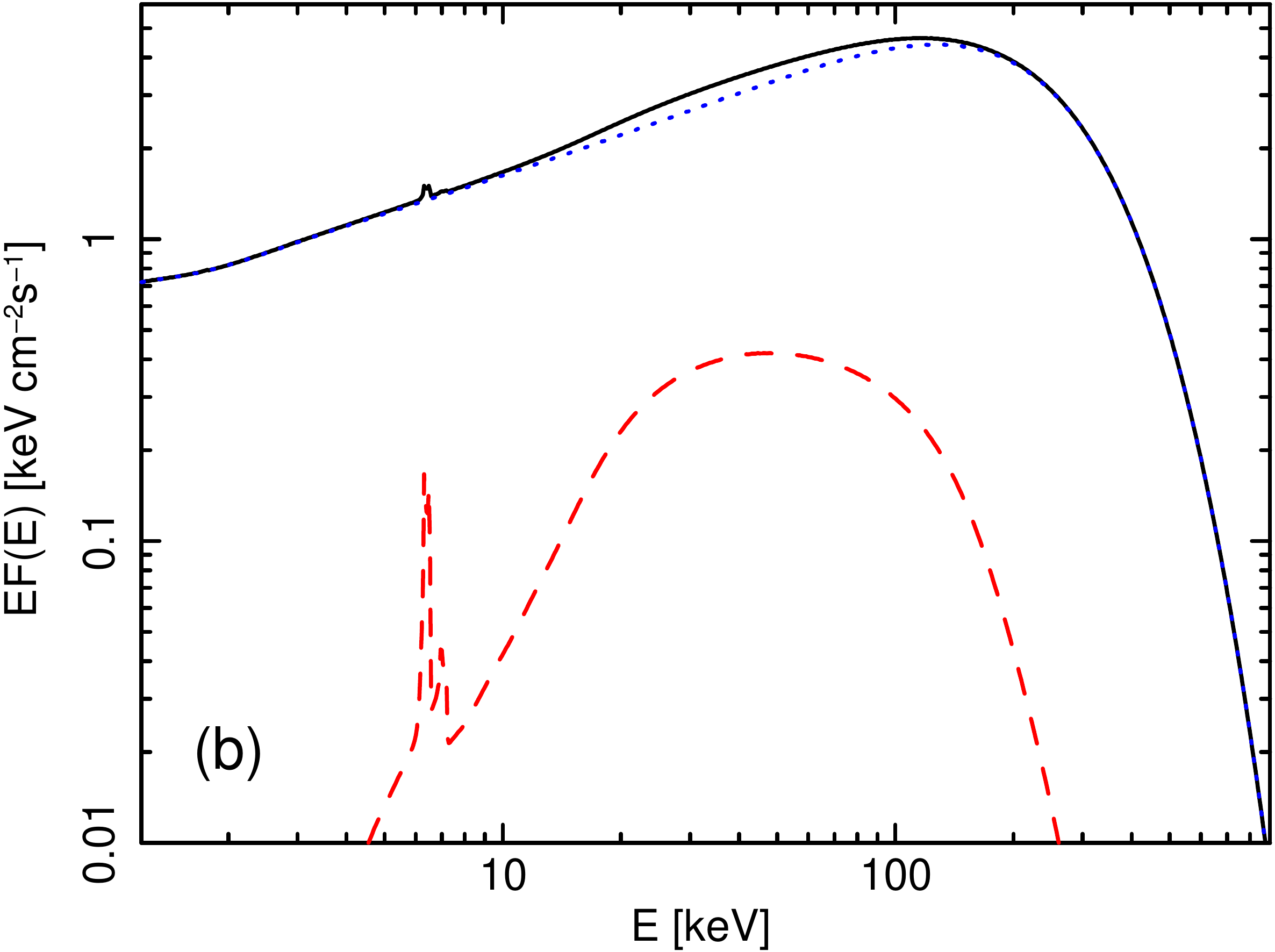}
  \caption{(a) The PCA spectrum fitted in the 4.1--10.1\,keV range with slightly relativistically broadened neutral reflection of a thermal Comptonization incident spectrum. The upper panel shows the unfolded spectra in black and the model in red. The bottom panel shows the data-to-model ratio including also the HEXTE data, where we see the presence of a weak excess below 4 keV and of additional Comptonization components above 10 keV. (b) The corresponding unabsorbed model. The blue dotted and red dashed curves show the incident thermal Comptonization and the reflected spectrum, respectively, and the solid curve is the sum.
}\label{4_10}
\end{figure}

While this model does not include the re-emission of absorbed photons, with the Fe K line added with a free normalization, equally good reflection fits can be obtained with either the reflection model {\tt xillver} or {\tt hreflect}. However, the best-fit break energy of the incident spectrum in previous fits hints at a soft excess component dominating below 3.8 keV. Therefore, we performed the fits with {\tt xillver} and {\tt hreflect} focusing on the energy range 4.1--10.1\,keV  (15 bins containing $5.4\times 10^7$ counts), thus excluding this component. Since neutral reflection contributes only weakly below 4 keV, this confirms that the Fe Ke line is added to the {\tt bexrav} model properly. We then consider models taking into account relativistic blurring (using either {\tt relxill} and {\tt reflkerr}), but since the data are very well fitted with static reflection, they give only a slight improvement of the fit. In particular, {\tt reflkerr} \citep{Niedzwiecki19} in the coronal geometry gives $R_{\rm in}\gtrsim 90 R_{\rm g}$ at $\chi_\nu^2 \approx 5.7/10$, see Table \ref{PCA_fit} and Figures \ref{4_10}a, b. Figure \ref{4_10}a shows the full energy range of 3--140\,keV, where we see a weak soft excess present below 4\,keV and an excess at $E\gtrsim 10$\,keV, which might be associated with inner, thus harder, Comptonization regions. 

\setlength{\tabcolsep}{3pt}
\begin{table}[t]
\centering            
\caption{The results of the spectral fit of the PCA spectrum in the 4.1--10.1 keV range (15 channels) with {\tt{jscrab*tbabs*reflkerr}}. }
\label{PCA_fit}      
\vskip -0.5cm                               
\begin{tabular}{ccccccc}
\hline 
$N_{\rm H}$ & $\Gamma$ & $R_{\rm{in}}$ & $\mathcal{R}$ & $N$& $\chi_\nu^2$ \\
$10^{22}{\rm cm}^{-2}$ &  & $\rg$ && $\frac{{\rm keV}}{{\rm cm}^2{\rm s}}$ & \\
\hline
$1.4_{-0.1}^{+0.2}$ & $1.58^{+0.01}_{-0.01}$ & $760_{-670}^{+\infty}$ & $0.12_{-0.01}^{+0.01}$ & 0.68& 6/10\\
\hline                    
\end{tabular}\\ 
\tablecomments{$i=30\degr$, $kT_{\rm e}=68$\,keV, $kT_{\rm bb}=0.2$\,keV, $\xi=1$\,erg\,cm\,s$^{-1}$ and $Z_{\rm Fe}=1$ are assumed, and $N$ gives the {\tt reflkerr} flux at 1 keV. The upper limit of infinity on $R_{\rm in}$ corresponds to the absence of the relativistic broadening. } 
\end{table}

\begin{figure}
  \centering
  \includegraphics[height=6.5cm,angle=-90]{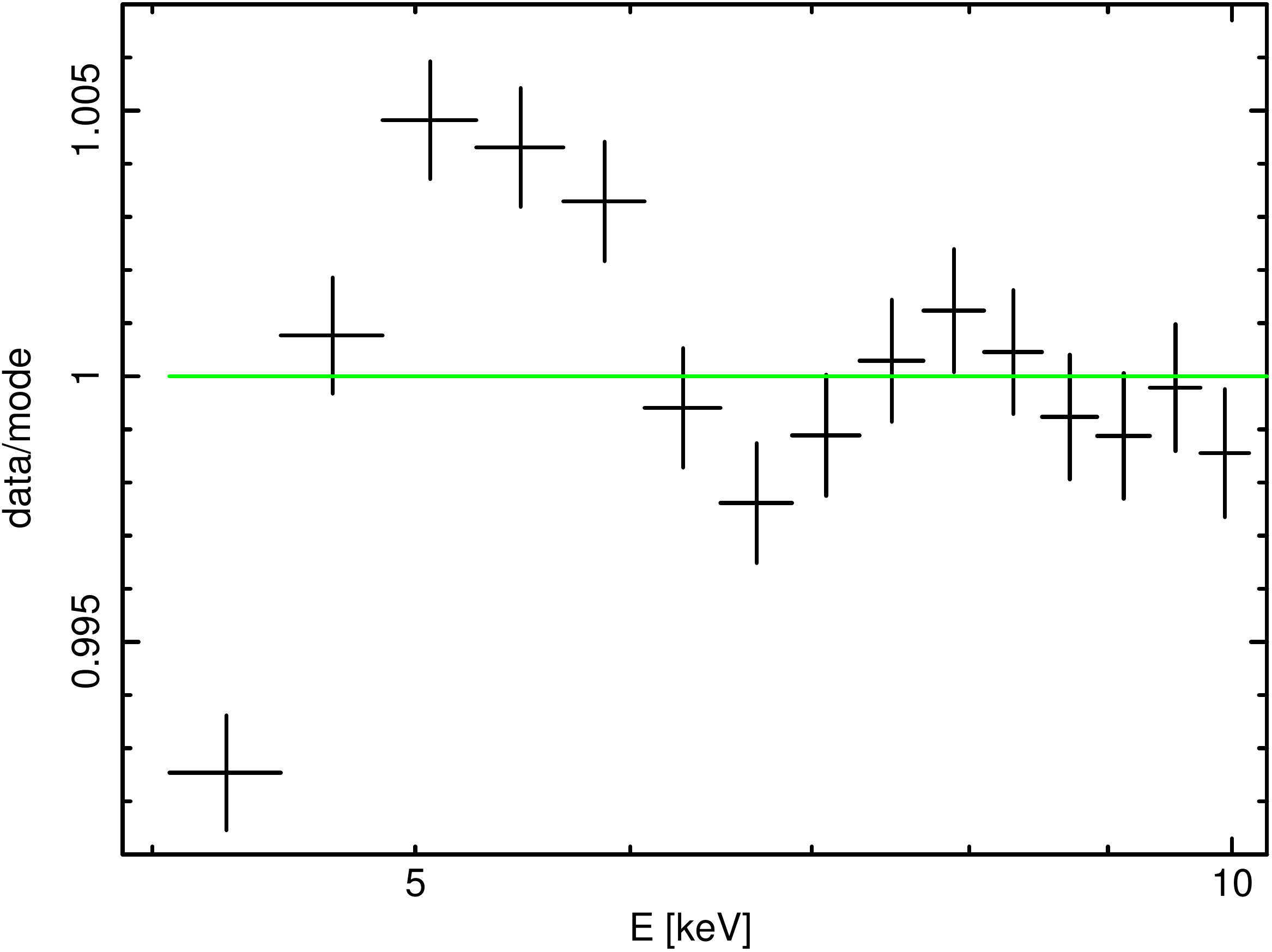}
  \caption{The ratio of the 4.1--10.1 keV PCA data to a fit with power law and a nearly neutral ($\xi=1$\,erg\,cm$^{-2}$\,s$^{-1}$) reflection absorbed by $N_{\rm H}=7\times 10^{21}$\,cm$^{-2}$.
}\label{ratio4_10}
\end{figure}

\begin{figure}
  \centering
  \includegraphics[height=6.5cm,angle=-90]{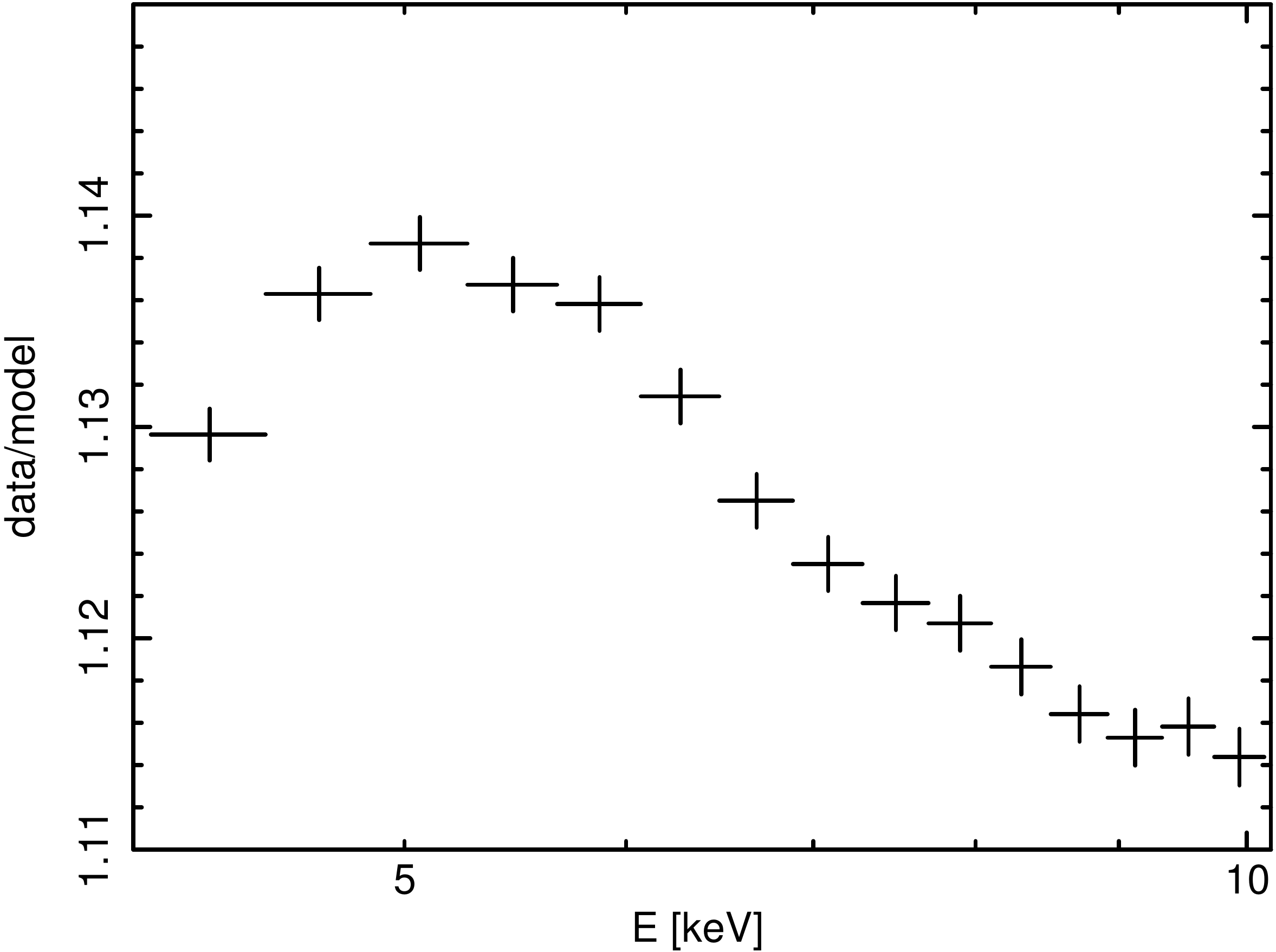}
  \includegraphics[height=6.5cm,angle=-90]{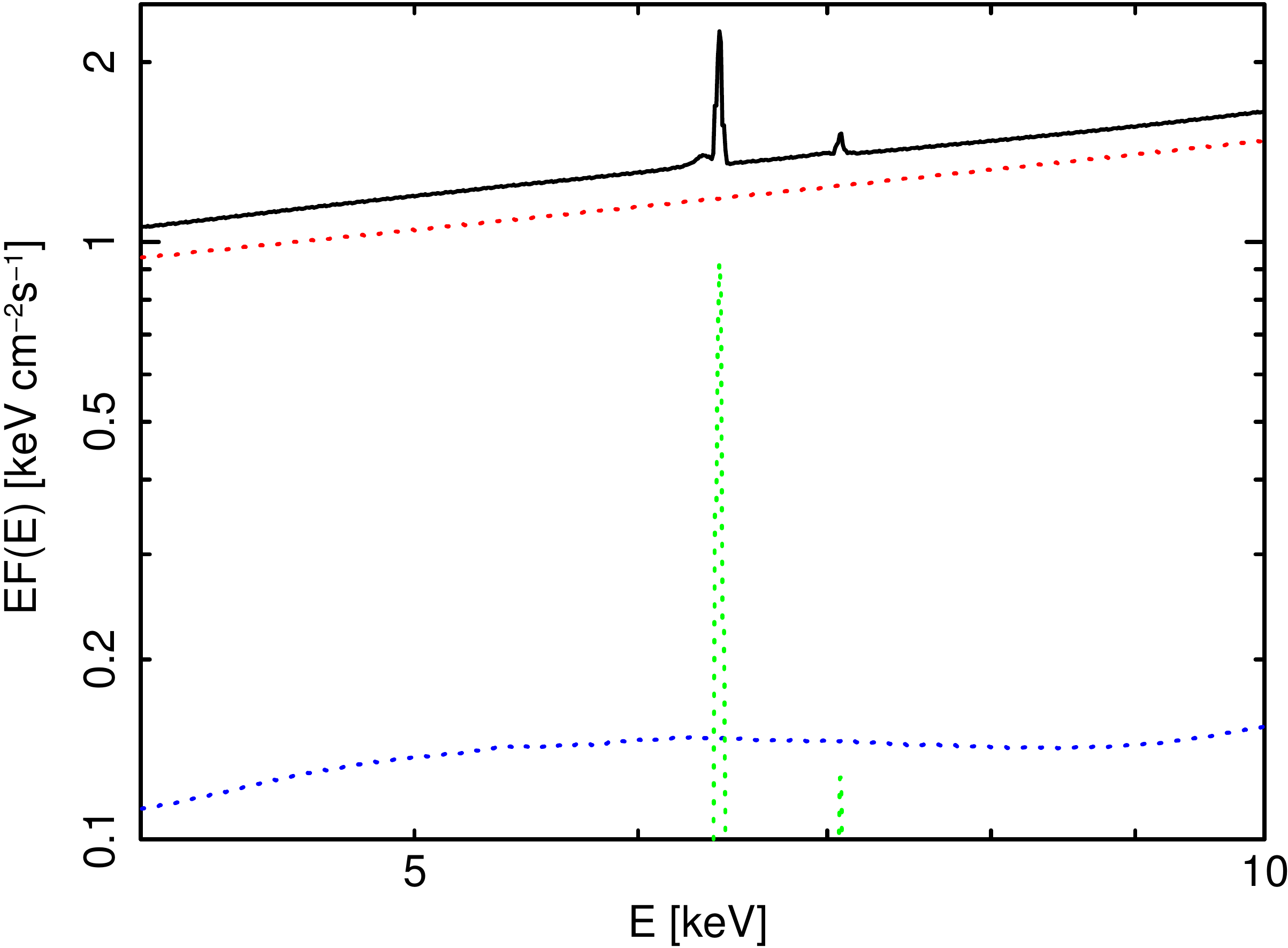}
  \caption{(a) The 4.1--10.1 keV PCA residuals (shown as ratios) seen after removing the relativistic reflection from the coronal model (\#4) to the XRT/PCA/HEXTE data (which yield $N_{\rm H}=7.0^{+0.3}_{-0.2}\times 10^{21}$\,cm$^{-2}$). (b) The components of the model in the same range. The red, blue and green curves show the incident continuum (a power law in this range), the relativistic reflection and the remote, static reflection, respectively. The black curve shows the sum. We see that the relativistic reflection is blurred so strongly that it effectively becomes another continuum component.
}\label{coronal_detail}
\end{figure}

Using the abundances of \citet{WAMC00} for the ISM absorption yields very similar results (except for higher best-fit values of $N_{\rm H}$). We have also tested a number of other simple models fitted to the 4--10 keV range, but none gave a reasonable fit. For example, an absorbed power law fit yields $\chi_\nu^2 \approx 264/12$, and an addition of a Gaussian line (narrow at 6.40 keV at the best fit) yields $\chi_\nu^2 \approx 25/9$. 

Still, an important problem with the models fitted to the $\leq$10.1\,keV range is that they require the presence of absorption significantly stronger than that fitted to the XRT data. E.g., the {\tt bexrav} model yields $N_{\rm H}\approx 1.33^{+0.20}_{-0.15}\times 10^{22}$\,cm$^{-2}$, and similar values are obtained in the 4.1--10.1\,keV models. This is twice as much as the XRT fit of $\approx 6.5^{+0.4}_{-0.3}\times 10^{21}$\,cm$^{-2}$. This could be due to either an intrinsic absorption occurring after the end of the \swift observations and dominating the average spectrum or a complexity of the intrinsic spectrum, e.g., the presence of another spectral component in addition to the nearly-static reflection. Figure \ref{ratio4_10} shows the ratio of the 4.1--10.1\,keV PCA data to a model with a power law and a static, nearly-neutral, reflection absorbed by $N_{\rm H}=7.0\times 10^{21}$\,cm$^{-2}$. We have chosen such a model since all of the models fitting the broad-band data in Section \ref{joint} include static reflection in addition to the relativistic one. The shown residuals can be interpreted as an additional weak Fe K edge and either a moderately broadened and redshifted Fe K line, or a hardening of the spectrum below $\sim$5\,keV (in addition to a broad Fe K edge). The large fitted value of $N_{\rm H}$ may also be an effect of fitting the emission of an inhomogeneous source by a single model. As we see in Figure \ref{4_10}a, the presence of a harder contribution is required above 10\,keV. However, it will also contribute below 10\,keV, implying that the actual 4--10\,keV spectrum is concave.

On the other hand, the broad-band fits in Sections \ref{joint}--\ref{quasi} show rather different components. Figure \ref{coronal_detail}a shows the data/model ratio resulting from removing the relativistic reflection from the coronal broad-band model, \#4 in Table \ref{models}. While we see a hardening below $\sim$5\,keV, we do not see any edge. Figure \ref{coronal_detail}b shows the model components, where we see that the relativistic reflection plays here the role of another continuum component, with hardening both below $\sim$5\,keV and above $\sim$9\,keV. Similar decomposition is seen in the other broad-band models of Section \ref{joint}. We note that the $\chi^2$ contributions to the 4.1--10.1\,keV of those models (Section \ref{joint}) is $\approx$15, while the very simple model of nonrelativistic reflection and absorption to the same range yields $\chi^2\approx 6$. Thus, while those models do provide relatively good overall description of the broad-band spectrum with $\chi^2_\nu\sim 1$, they do not account well for the details of the 4--10\,keV range, which contains the Fe K line and edge, crucial for reflection.

Interpretation of these results is not obvious. It is highly remarkable that such simple reflection models fit the data at $\leq$10.1\,keV very well. But if the absorption column is indeed as measured by the XRT, these models require that the incident continuum is curved in a way emulating the extra absorption. On the other hand, highly relativistic models fit relatively well the full broad-band data. However, their parameters are clearly unphysical, as discussed in Sections \ref{joint}--\ref{quasi}.

\section{Mutual Connection of Disk Blackbody, Comptonization and Reprocessing}
\label{mutual}
\subsection{The Formalism}
\label{formalism}

We discuss here a way in which the disk blackbody, reflection and primary (Comptonization) spectral components can be connected in a self-consistent way. Namely, a fraction, $f_{\rm sc}$, of the blackbody emission (with the luminosity $L_{\rm disk}$) is Compton upscattered to form the primary component with the luminosity $L_{\rm C}$ (though a part of the Comptonization component may be from upscattering of synchrotron photons). The same fraction of the reflection spectrum is Comptonized in the hot medium, which main effect is to reduce the observed normalization of that spectrum by a factor $1-f_{\rm sc}$. Thus, the actual reflection relative fraction can be estimated as ${\cal R}_0={\cal R}/(1-f_{\rm sc})$.  

The reflection component originates from irradiation of the cold medium by the primary radiation. That reflection and the associated emission line spectrum, fitted to the X-ray data, contain only a fraction of the irradiating flux. The remainder of it is absorbed and reemitted as a reprocessed spectrum, roughly resembling a disk blackbody when integrated over the disk surface \citep{ZDM20}. The relative reflection fraction, ${\cal R}_0$, and the backscattering albedo, $a$ ($\approx 0.3$--0.7 for typical LHS spectra; \citealt{ZDM20}), determine the amount of the quasi-thermal re-emission contributing to the disk blackbody. This contribution can be approximately estimated from the total irradiating luminosity of 
\begin{equation}
L_{\rm irr}\approx {\cal R} L_{\rm C}/(1-f_{\rm sc}),\quad L_{\rm C}=4\pi d^2 F_{\rm C},\label{Lirr}
\end{equation}
where $F_{\rm C}$ is the observed flux in the Comptonization component. Note that $L_{\rm C}$ estimated here from $F_{\rm C}$ is not necessarily the total luminosity emitted by the Comptonizing plasma. In a disk-corona geometry, $L_{\rm C}$ is is only the luminosity emitted outward. The remaining contribution to the disk blackbody emission is from the internal dissipation. The contribution to the disk blackbody emission at the disk inner edge due to irradiation has the color temperature $T_{\rm in,irr}$, which is given by
\begin{eqnarray}
&\sigma \left(\frac{T_{\rm in,irr}}{\kappa}\right)^4\approx \frac{(1-a)L_{\rm irr}}{2\pi R_{\rm in}^2}\approx \frac{2(1-a){\cal R} d^2 F_{\rm C}}{R_{\rm in}^2 (1-f_{\rm sc})}, \label{Tirr}\\
&T_{\rm in}^4=T_{\rm in,irr}^4+T_{\rm in,diss}^4,\label{Tsum}
\end{eqnarray}
where $T_{\rm in}$ is the inner color temperature fitted in the disk blackbody component, $T_{\rm in,diss}$ is that due to the intrinsic dissipation, $\kappa$ is the color correction, and an irradiation profile $\propto R^{-3}$ (as in the disk blackbody) is assumed to relate the irradiating flux to $L_{\rm irr}$ \citep{ZDM20}. Note that since $T_{\rm in,irr}\leq T_{\rm in}$ is required, this gives a lower limit on $R_{\rm in}/d$ based on the value of $T_{\rm in}$ and independently of the fitted disk blackbody normalization.

For the blackbody emission of the disk, we use here the {\tt diskbb} model, which does not include the zero-stress boundary condition at the inner radius. Its dimensionless normalization, $N_{\tt diskbb}$, is related to the disk inner radius as
\begin{equation}
R_{\rm in}=x\kappa^2 d N_{\tt diskbb}^{1/2}\cos^{-1/2}\!i,
\label{rin_disk}
\end{equation}
where $x\equiv 10^4{\rm cm}/1\,{\rm kpc}\approx 3.24\times 10^{-18}$. Note that this gives a constraint on $R_{\rm in}/d$ based on the normalization and $i$ only, and independent of the fitted $kT_{\rm in}$. This constraint can be compared to $R_{\rm in}/R_{\rm g}$ from reflection spectral fitting. 

Equation (\ref{rin_disk}) can be inserted in Equation (\ref{Tirr}), yielding
\begin{equation}
\sigma T_{\rm in,irr}^4\approx \frac{2(1-a){\cal R} F_{\rm C}\cos i}{x^2  (1-f_{\rm sc})N_{\tt diskbb}}. \label{Tirr2}
\end{equation}
Notably, this estimate is independent of $d$, $M$ and $\kappa$. It then constrains $T_{\rm in,diss}$ via Equation (\ref{Tsum}) and since $T_{\rm in,irr}\leq T_{\rm in}$, it serves as a self-consistency check of a fitted model. 

The observed energy-integrated flux and the luminosity in the disk blackbody are
\begin{eqnarray}
&F_{\rm disk}=2\left(\frac{R_{\rm in}}{d}\right)^2\sigma \left(\frac{T_{\rm in}}{\kappa}\right)^4\cos i=2 x^2 N_{\tt diskbb} \sigma T_{\rm in}^4,
\label{fdisk}\\
&L_{\rm disk}=\frac{2\pi d^2 F_{\rm disk}}{\cos i}=4\pi \sigma T_{\rm in}^4 x^2 d^2 \frac{N_{\tt diskbb}}{\cos i},
\label{ldisk}
\end{eqnarray}
respectively. The mass accretion rate is\footnote{Note that if the zero-stress inner boundary term were included in the disk blackbody model, a factor of 3 would appear in the denominator.}
\begin{equation}
\dot M=\frac{8\pi R_{\rm in}^3\sigma T_{\rm in,diss}^4}{(1-f_{\rm c})\kappa^4 G M},
\label{mdot}
\end{equation}
where $T_{\rm in,diss}$ is given by Equations (\ref{Tsum}) and (\ref{Tirr2}). Then, $\dot M$ is related to the bolometric luminosity,
\begin{eqnarray}
&L_{\rm bol}\approx 4\pi d^2 F_{\rm C}+2\pi d^2 \frac{F_{\rm disk}}{\cos i}=\nonumber\\
&=4\pi d^2 F_{\rm C}+4\pi D^2 \sigma T_{\rm in}^4 x^2 \frac{N_{\tt diskbb}}{\cos i} =\epsilon \dot M c^2,
\label{Lbol}
\end{eqnarray}
where $\epsilon$ is the accretion efficiency.

Five quantities obtained (with uncertainties) from fitting a given set of observations are $F_{\rm C}$, ${\cal R}$, $f_{\rm sc}$, $T_{\rm in}$ and $N_{\tt diskbb}$. Then, the quantities estimated within some either theoretical or observational uncertainty ranges are $a$, $\kappa$, $\epsilon$, $i$, $d$, $f_{\rm c}$, while $\dot M$ is an inferred source parameter. The above equations allow us to constrain the source parameters and check their self-consistency.

Then, the fitted values of ${\cal R}$ and $f_{\rm sc}$ constrain the flow geometry. The effective solid angle that the disk subtends as seen by the hot flow is $2\pi{\cal R}_0$, and the solid angle that the Comptonizing region subtends as seen by the disk is $2\pi f_{\rm sc}$. Given the inhomogeneity of the source, those quantities correspond to average values describing the accretion flow.

\subsection{Constraints on the Hard State of \source}
\label{constraints}		

We use the value of ${\cal R}\approx 0.12^{+0.01}_{-0.01}$ from Table \ref{PCA_fit}, $f_{\rm sc}\approx 0.24^{+0.13}_{-0.11}$, $kT_{\rm in}\approx 0.11\pm 0.02$\,keV, $N_{\tt diskbb}\approx 0.9^{+2.8}_{-0.5}\times 10^6$ from Table \ref{XRT_fit}, and $F_{\rm C}\approx 2.0\times 10^{-8}$\,erg\,cm$^{-2}$\,s$^{-1}$, which is the Comptonization flux of the broad-band XRT/PCA/HEXTE model shown in Figure \ref{coronal}. However, we set the lower limit on $f_{\rm sc}$ equal to zero, accounting for uncertainties in the geometry and the effect of Compton scattering on the reflection spectrum. Then, the above value of ${\cal R}$ is to be considered tentative given the caveat regarding the value of $N_{\rm H}$ given in Table \ref{PCA_fit}; however, that ${\cal R}$ is the lowest of all of the considered models, which is conservative choice, minimizing the efffect of the thermal re-emission of a part of the irradiating flux. We assume $i=30\degr$, but the results depend relatively weakly on that within the likely range of $\lesssim\!60\degr$. We consider $\kappa=1.5\pm 0.2$ \citep{Davis05}, $a=0.5\pm 0.2$ \citep{ZDM20}, $M=10\pm 5\msun$ and $d=6\pm 2$\,kpc as likely allowed ranges. We calculate the uncertainty ranges of the derived parameters from the extrema of the above parameters, which is a conservative approach, yielding larger uncertainties than those from the propagation of errors.

We first consider constraints on the radius. Our fits yield three independent constraints,
\begin{eqnarray}
&\frac{R_{\rm in}}{R_{\rm g}}\gtrsim 90,
\label{R_reflection}\\
&\frac{R_{\rm in}}{R_{\rm g}}\approx 100_{-40}^{+100}\left(\frac{\kappa}{1.5}\right)^2 \frac{d}{6\,{\rm kpc}} \frac{10\msun}{M},
\label{R_N}\\
&\frac{R_{\rm in}}{R_{\rm g}}\gtrsim 130_{-30}^{+50} \left(\frac{1\!-\!a}{0.5}\right)^{1/2}\left(\frac{\kappa}{1.5}\right)^2  \frac{d}{6\,{\rm kpc}} \frac{10\msun}{M}. \label{R_T}
\end{eqnarray}
The first one, Equation (\ref{R_reflection}), follows from the reflection fit to the 4.1--10.1\,keV PCA data, Table \ref{PCA_fit}. While it should be considered tentative, it agrees relatively well with the other two. The next constraint, Equation (\ref{R_N}), follows solely from the normalization of the disk blackbody fit, Table \ref{XRT_fit}, and it yields $R_{\rm in}/R_{\rm g}\gtrsim 20$ for $\kappa\gtrsim 1.3$, $d\gtrsim 4$\,kpc, $M\lesssim 15\msun$. The final constraint, Equation (\ref{R_T}), is due to the irradiation, see Equations (\ref{Tirr}--\ref{Tsum}), and is determined by the reflection fraction, the irradiating flux, and the inner temperature and the scattered fraction of the disk blackbody. 

Then, Equation (\ref{Tirr2}) yields a constraint on $kT_{\rm in,irr}$ independent of $d$, $M$ and $\kappa$.  We obtain $kT_{\rm in,irr}\approx 0.13_{-0.06}^{+0.05}$\,keV, while we require it to be $\leq kT_{\rm in}\approx 0.11\pm 0.02$\,keV. Within these uncertainties, no constraint on $f_{\rm c}$ is found. Independent of that value, the irradiation requires a large truncation radius, $R_{\rm in}/R_{\rm g}\gtrsim 30$, as implied by Equation (\ref{R_T}) for $\kappa\gtrsim 1.3$, $d\gtrsim 4$\,kpc, $M\lesssim 15\msun$. 

We then consider the constraints on $\dot M$. The bolometric luminosity from Equation (\ref{Lbol}) is $L_{\rm bol}\approx 9.5_{-0.4}^{+2.2}(d/6\,{\rm kpc})^2 10^{37}$\,erg\,s$^{-1}$. This corresponds to (0.06--$0.08)(d/6\,{\rm kpc})^2 (M/10\msun)^{-1} \ledd$ (at the H fraction of $X=0.7$), and to the actual accretion rate of 
\begin{equation}
\dot M\approx (1.0\!-\!\!1.3) (d/6\,{\rm kpc})^2(\epsilon/0.1)^{-1} 10^{18}{\rm g}\,{\rm s}^{-1}.
\label{mdot2}
\end{equation}
On the other hand, the maximum $\dot M$ from Equation (\ref{mdot}) at the maximum possible $kT_{\rm in,diss}\approx 0.13$\,keV is $2.7(\kappa/1.5)^2 (d/6\,{\rm kpc})^3 (M/10\msun)^{-1} 10^{19}$\,g\,s$^{-1}$. This, in principle, allows for $\epsilon\ll 0.1$. However, given that the PCA count rate in this LHS is quite close (\garcia) to that of the following soft state, where we expect $\epsilon\gtrsim 0.1$ and $\dot M$ higher than in the LHS, values of $\epsilon\ll 0.1$ in the LHS are ruled out. Thus, this luminous LHS accretion flow appears to have $\epsilon\sim 0.1$, in agreement with the theoretical prediction of \citet{YN14}, see their fig.\ 2.

These considerations show that a large disk truncation radius in \source is implied by three independent constraints. One is given by the tentative result of the fit to the PCA spectrum at $\leq$10\,keV (Section \ref{inhomogeneity}), indicating that the reflection component present in \source is only weakly blurred by relativistic effects. This estimate of the truncation radius is then in agreement with the XRT data, which were not studied in \garcia. Indeed, the normalization of the disk blackbody model fitted to the XRT data yields $R_{\rm in}/R_{\rm g}\gtrsim 20$ (Equation \ref{R_N}). Moreover, the fitted inner temperature of the same model is so low that irradiation of the disk by the primary source has to be weak, requiring $R_{\rm in}/R_{\rm g}\gtrsim 30$ (Equation \ref{R_T}), which constraint is independent of the previous one. Thus, all three constraints taken together rule out models with the disc extending to the vicinity of the ISCO.

\section{Summary and Discussion}
\label{discussion}

In Section \ref{S:XRT}, we fitted the 0.55--6\,keV spectrum of \source from the XRT covering the first 9 days of the \xte observations. This spectrum can be fitted by a disk blackbody and either a power law or thermal Comptonization. In either case, we obtained a large normalization of the disk blackbody with, $N_{\rm diskbb}\sim 10^6$, and a very low inner temperature of $\approx$0.1\,keV. This normalization implies a large disk truncation radius, $R_{\rm in}\gtrsim 20 R_{\rm g}$ (where the limit depends on the uncertain color correction, BH mass and the distance, see Section \ref{constraints}). The obtained temperature is one of the lowest measured in luminous hard states of accreting BH binaries. 

We then studied the 3--140\,keV spectra from \xte, first those data alone and then jointly with the XRT data (Section \ref{joint}). We can find well-fitting models in either case. The models require the disk to extend relatively close to the BH, $R_{\rm in}\sim 3 R_{\rm ISCO}$, i.e., much less than the $R_{\rm in}$ required by the XRT fits. However, all of the lamppost models have the luminosity measured in the lamppost frame exceeding the threshold for e$^\pm$ pair equilibirum by factors $>10^2$, leading to runaway pair production. In addition, most of the emitted photons are captured by the BH, implying the mass accretion rate in that hard state exceeding that in the following soft state, which is unlikely. Also, the observed flux in the disk-reflected photons is significantly lower than the one implied by the lamppost geometry. A possible solution to this problem is the presence of a gravitationally focused emission from the bottom lamp; this, however, requires the inclination to be in a narrow range around $10\degr$. We also tested a coronal-like model with a power-law irradiation profile, finding the best fit with an extremely steep index of $q\approx 15.0^{+1.9}_{-1.4}$, which is not physical. Thus, while we can find phenomenological models of the broad-band spectrum, they violate a number of physical constraints.

Then, we considered the effects of the intrinsic disk dissipation and quasi-thermal re-emission of the absorbed part of the irradiating flux. To account for that, we have developed a new {\sc xspec} lamppost model, \texttt{reflkerr\_lpbb}, described in Appendix \ref{thermal}. We found that imposing a correct value of the backscattering albedo prevents a satisfactory fit to the joint data, as well as the obtained model with a high $\chi^2_\nu$ violates the pair equilibrium in an extreme way and requires most of the emitted photons to be captured by the BH. 

These results strongly suggest that the X-ray source in \source is inhomogeneous, in agreement with the independent timing results obtained by \munoz, and in agreement with recent results for other accreting BH binaries (e.g., \citealt{Mahmoud19}). We considered this possibility in Section \ref{inhomogeneity}. We found that the spectrum from the PCA at $\leq$10.1\,keV (containing 64\% of the total of $10^8$ counts) is very well fitted with almost static Compton reflection, with the disk inner radius constrained to $R_{\rm in}\gtrsim 100\rg$. The reflecting medium is weakly ionized and at the Fe abundance close to the solar value. In particular, the $\chi^2$ of the best fit to the 4.1--10.1\,keV range is three times lower than the $\chi^2$ contributions to that band of the best-fit broad-band models of Section \ref{joint}. This energy range is crucial to precisely constrain the relativistic broadening since it contains the intrinsically sharp Fe K lines and edges. The $\leq$10.1\,keV fit underpredicts the observed spectrum at higher energies up to the maximum of 10\%. This appears to be due to the inhomogeneity of the accretion flow, which emission becomes harder with the decreasing radius, as evidenced, e.g., by the observed hard X-ray lags. However, our models for the energy range $\leq$10.1\,keV require the absorbing column density to be about twice the value determined from the XRT fit. This is an important caveat, to which we have found no qualitative solution. 

Our finding that the apparent relativistic effects strongly increase when spectra at $>$10\,keV are included in fits appears to be due to the upward spectral curvature at those energies. This results in either the fitted reflection being stronger and/or the fitted incident continuum being harder than those implied by the data at $\leq$10\,keV. In the case of the broad-band fits of Section \ref{joint}--\ref{quasi}, both of these effects are present. Since the $\leq$10\,keV data show only weak Fe K line and edge, the stronger reflection component implied by the upturn at 10 keV when fitted as a reflection hump needs to be heavily blurred in order to resemble another continuum component. Then, the included remote reflection accounts for the narrow Fe K line present in the data. This is clearly seen in Figure \ref{coronal_detail}b, where we see that the reflection spectrum in one of those fits is virtually featurless. Also, it is significantly stronger than the reflection implied by the fit for $\leq$10\,keV. We argue this is a signature of a more complex, spectrally inhomogeneous, primary continuum, as also suggested by timing results of these and other sources.

We note that \citet{Nowak11} found that the spectra of Cyg X-1 also show spectral upturns above 10\,keV stronger than those expected from the presence of their fitted reflection. Then, the effect we have found may explain the discrepancy between the results of \citet{Basak16}, who fitted the \xmm spectra of the LHS of GX 339--4 below 10 keV finding only weak relativistic broadening of reflection, and those of \citet{Garcia15,Wang-Ji18}, who fitted broad-band data finding much stronger relativistic effects. However, while we have found a good agreement between the truncation radii from both the normalization of the disk blackbody and reflection fits, \citet{Basak16} found that imposing those two fit radii to be equal led to an unlikely behavior of the inner radius (decreasing with the increasing spectral hardness) for seven LHS observations of GX 339--4. As we have pointed out, the disk blackbody model remains approximately valid in the presence of irradiation, which only increases the inner temperature with respect to the case with intrinsic disk dissipation only. Thus, irradiation alone does not explain that result. It may be explained by the presence of a soft X-ray excess in addition to the disk blackbody in GX 339--4 with the fitted model not including the former. On the other hand, the XRT data for \source do not show any soft excess. 

The spectrum irradiating the disk implied by the 4.1--10.1\,keV fits when extrapolated to higher energies accounts for only $\approx$90\% of the spectrum at $\sim$50\,keV, thus showing that an additional, harder, spectral component is required. Without detailed studies of the timing properties of \source (which is beyond the scope of this paper), the spectral shapes of the two components cannot be reliably constrained. 

In Section \ref{mutual}, we developed a formalism to self-consistently connect the parameters of the disk blackbody, Comptonization, reflection/reprocessing and the mass accretion rate. Based on it, we have found that both the disk blackbody normalization and the bolometric flux of the incident spectrum when constrained to yield the irradiation temperature of $\approx$0.1\,keV imply $R_{\rm in}\gtrsim 20 R_{\rm g}$. Thus, the strong truncation is implied in two independent ways and in agreement with the result for the spectrum at energies $\leq$10.1\,keV. 

The disk at $R>R_{\rm in}\gtrsim 20 R_{\rm g}$ and the hot flow only at $R<R_{\rm in}$ would imply small values of ${\cal R}$ and $f_{\rm sc}$. The moderate values found from the fits imply that there is an overlap between the disk and the hot flow, allowing then their stronger mutual interaction \citep{PVZ18}. 

The methods developed in this work can also be applied to the LHS of other accreting BH binaries. The evidence for the disk being close to the ISCO at $\gtrsim$1\% of $L_{\rm Edd}$ appears to be exclusively based on results of broad-band spectral fitting. It would be highly interesting to see whether constraining the fitted energy band to $\lesssim$10\,keV systematically leads to an increase of the truncation radius.

An interesting issue is also possible inhomogeneity of the X-ray sources in AGNs, in particular in radio-quiet Seyfert galaxies, which have X-ray spectra often similar to those in the LHS of accreting BH binaries. The approach developed in this work can be applied to those sources.

\section{Main Conclusions}
\label{conclusions}

Broad band fits assuming a single Comptonization region imply small truncation radii, but yield unphysical results from other points of view. The irradiation model developed by us, which still assumes a single Comptonization region, yields unsatisfactory fits. On the other hand, our analysis of the 4--10 keV spectrum suggests an inhomogeneity of the X-ray source. We conclude that such a solution may solve these discrepancies.

\section*{Acknowledgments}
We thank Javier Garc{\'{\i}}a for providing us with the average \xte spectra and for comments as the referee of the original version of this paper. We have benefited from discussions during Team Meetings of the International Space Science Institute (Bern), whose support we acknowledge. We also acknowledge support from the Polish National Science Centre under the grants 2015/18/A/ST9/00746 and 2019/35/B/ST9/03944, and the European Union's Horizon 2020 research and innovation program under the Marie Sk{\l}odowska-Curie grant agreement No.\ 798726.

\appendix
\section{Lamppost with quasi-thermal emission}
\label{thermal}

Our new lamppost model, \texttt{reflkerr\_lpbb}\footnote{\url{http://users.camk.edu.pl/mitsza/reflkerr}}, extends the \texttt{reflkerr\_lp} model \citep{Niedzwiecki19} by accounting for two effects. One is the re-emission of the irradiating flux absorbed by the disk taking place as quasi-thermal emission satisfying the Stefan-Boltzmann law. The other is the intrinsic disk emission, due to the viscous dissipation. 

We consider two X-ray sources (lamps) symmetrically located at the height $H$ at each side of the disk truncated at $R_{\rm in}$. The luminosity of each lamp measured at infinity is $L_{\rm l}$. We assume that the lamps are cooled by thermal Comptonization and the spectrum of the emitted radiation is described by \texttt{compps} (with $T_{\rm e}$, $T_{\rm seed}$, $\Gamma$ as free parameters of the spectrum) for the spherical geometry, so their intrinsic emission is isotropic. We use the GR transfer functions of our \texttt{reflkerr\_lp} model to find the directly observed flux, $F_{\rm obs}$, and the flux irradiating the disk at the distance $R$, $F_{\rm irr}(R)$. The local reflection spectrum for a given $F_{\rm irr}$ is given by the {\tt hreflect} model, which uses either {\tt xillver} or {\tt xillverCp} at $E\lesssim 10$\,keV, which assume the reflector density of $n= 10^{15}$\,cm$^{-3}$. The above features are identical to those in \texttt{reflkerr\_lp}, except that in order to determine the amplitude of the irradiating flux we need to specify the distance to the source, $d$, and the BH mass, $M$, which are therefore free parameters of \texttt{reflkerr\_lpbb}. 

As follows from Equation (\ref{xi}), $F_{\rm irr}$ at typical values of the ionization parameter of $\xi\sim 10^{3}$\,erg\,cm\,s$^{-1}$ has values characteristic to active galactic nuclei, but much lower than those expected in accreting BH binaries. Therefore, a major part of the absorbed flux in the model is re-emitted below 0.1\,keV (which is the minimum energy at which {\tt xillver} spectra are calculated in {\sc xspec}), whereas it is, in reality, re-emitted at higher energies in BH binaries (e.g., \citealt{ZDM20}). In order to account for this, we assume that the part of the irradiating flux falling below 0.1\,keV in {\tt xillver} is thermalized and re-emitted as a local diluted blackbody radiation with the effective temperature given by Equation (\ref{teff}). (This is similar to the approach of \citealt{PVZ18} except that they used a fixed blackbody temperature of 50\,eV.) Here, we call the fraction of the incident flux re-emitted above 0.1\,keV the {\sl albedo parameter}, $a$. This is larger than the exact albedo (which corresponds to the backscattering only) since spectra from {\tt xillver, xillverCp} at $\xi\sim 10^{3}$\,erg\,cm\,s$^{-1}$ and $\Gamma<2$ are dominated by backscattering at $E\gtrsim 0.2$--1\,keV only (see \citealt{Garcia16}) as well as they contain the Fe K complex at $\approx$6--7\,keV. However, this approach is conservative in the sense it {\it underestimates\/} the average energy at which the absorbed flux is re-emitted, since the actual re-emitted spectra shift to higher energies with the increase of $n$ (and thus with the corresponding increase of $F_{\rm irr}$; \citealt{Garcia16}). Therefore, departures from models neglecting re-emission, e.g., \texttt{reflkerr\_lp}, are minimized. Further, we can achieve the conservation of energy. The model has $a$ as a free parameter. After fitting, we can check whether the used value was correct. For that, we need to average the {\tt hreflect} spectra for the best-fit parameters over the viewing angle, in order to determine the total luminosity emitted by the disk. 

Similarly as in \texttt{reflkerr\_lp}, we allow to scale $F_{\rm irr}$ by $\mathcal{R}$, in which case both the reflected and the thermally re-emitted components are self-consistently scaled by the same factor. However, the results presented in this paper correspond to the actual values predicted for the lamppost geometry, i.e.\ $\mathcal{R}=1$. 

In \texttt{reflkerr\_lpbb}, $L_{\rm l}$ is determined by the total observed flux in the Comptonization component, corrected by the light-bending factor (i.e.\ light-bending reduction of the direct radiation from the top lamp and enhancement of radiation from the bottom lamp, if it is visible), and the distance, $d$. We determine the accretion rate, $\dot M$, at which the accretion flow is able to power the two lamps, using the relation $(f_{\rm c} \epsilon_1 + \epsilon_2) \dot M c^2 = 2 L_{\rm l}$, where $\epsilon_1$ and $\epsilon_2$ are the efficiencies of the accretion flow at $R > R_{\rm in}$ and $R < R_{\rm in}$, respectively, and the fraction of the power dissipated at $R > R_{\rm in}$ which is transfered outside the disk, $f_{\rm c}$, is a free parameter. Following \citet{Thorne74}, we find the accretion efficiencies, $\epsilon_1 = 1 - \mathcal{E}(R_{\rm in})/(mc^2)$ and $\epsilon_2 = [\mathcal{E}(R_{\rm in}) - \mathcal{E}(R_{\rm ISCO})]/(mc^2)$, where $\mathcal{E}(R)$ is the energy of a particle with the rest-mass $m$ at the circular orbit at $R$. For $R_{\rm in} = R_{\rm ISCO}$ we have $\epsilon_2 = 0$, whereas $\epsilon_1$ is the efficiency of an untruncated disk, e.g., $\epsilon_1\approx 0.057$ for $a=0$ and $\epsilon_1\approx 0.32$ for $a=0.998$. 

We use the formula of \citet{Page74} for the local energy release per unit area of the disk accreting at $\dot M$, $F_{\rm diss}$. The effective temperature of the sum of the thermally radiated part of the locally dissipated energy and the thermalized fraction of the locally irradiating flux is given by Equation (\ref{teff}). For illustration purposes, we also define $\sigma T^4_{\rm eff,diss} = (1 - f_{\rm c})F_{\rm diss}$ and $\sigma T^4_{\rm eff,irr} = (1 - a)F_{\rm irr}$, cf.\ Equation (\ref{Tsum}).

\begin{figure}
\centerline{\includegraphics[height=5.5cm]{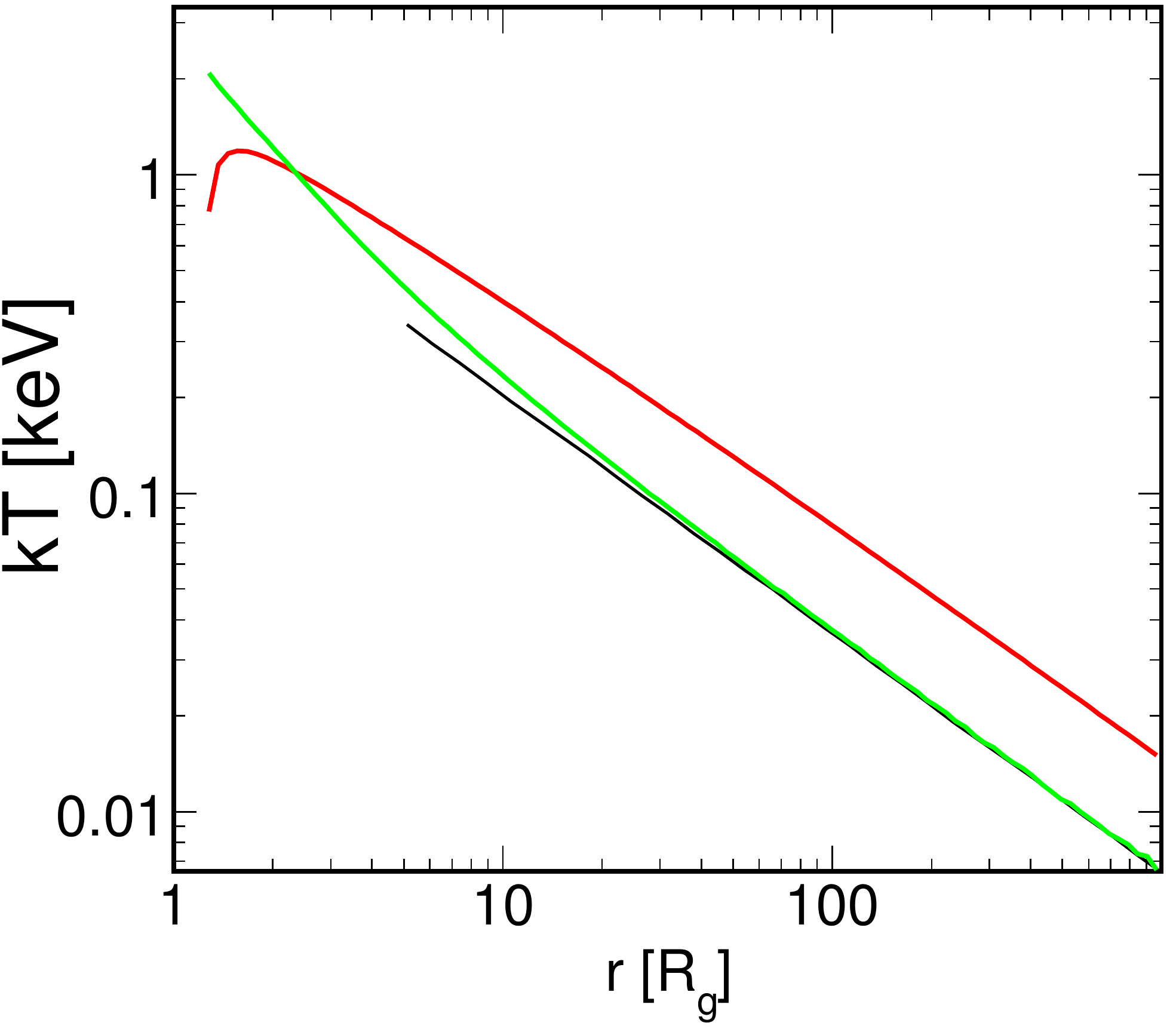}}
\centerline{\includegraphics[height=5.5cm]{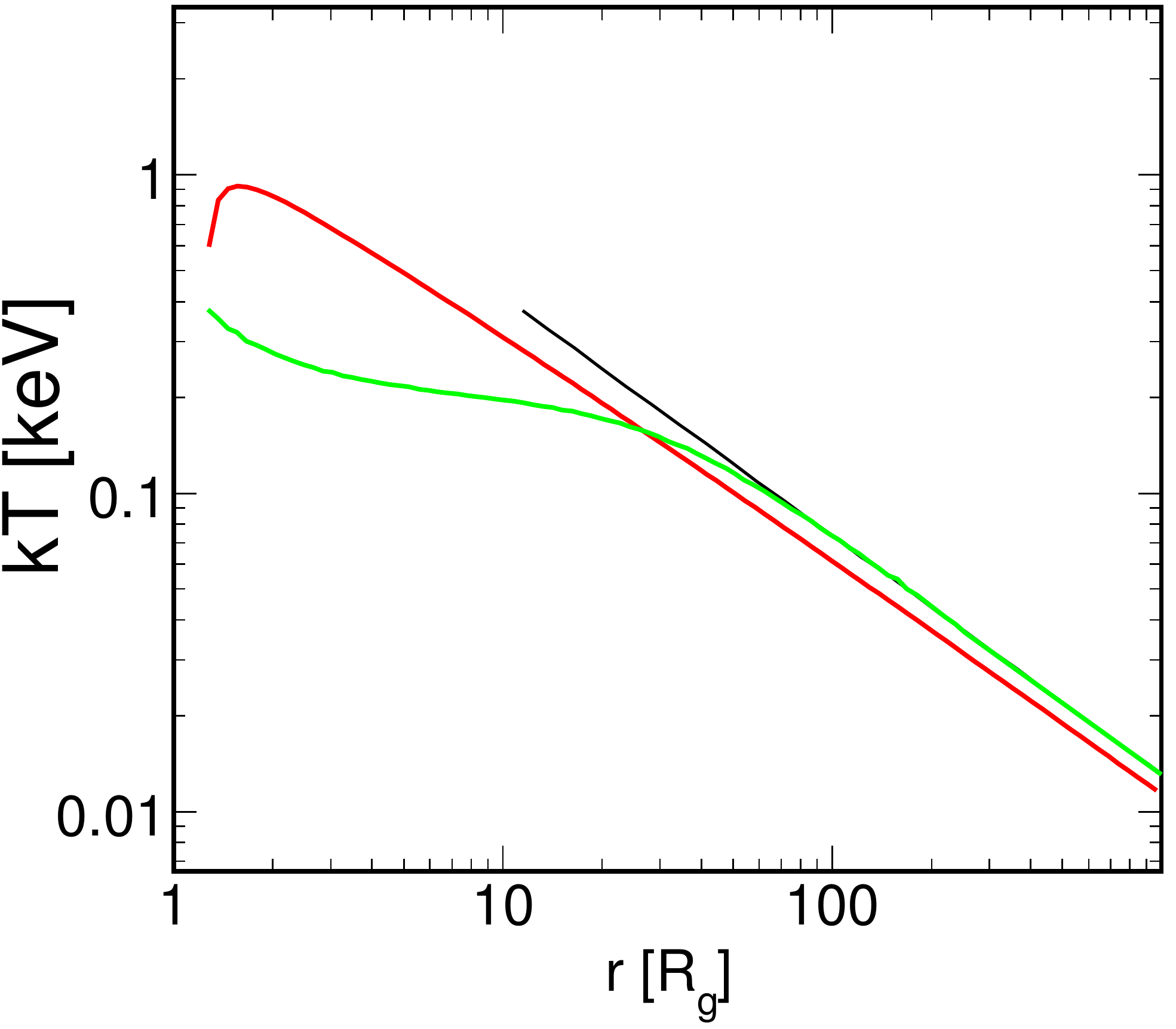}}
\caption{Temperature profiles for a disk around the black-hole with $a_*=0.998$ and $M = 9.6 \msun$, illuminated by a source with (top) $L_{\rm l} = 1.6 \times 10^{38}$\,erg\,s$^{-1}$ at $H=2 R_{\rm g}$ and (bottom) $L_{\rm l} = 5.7 \times 10^{37}$\,erg\,s$^{-1}$ at $H=30 R_{\rm g}$. The red curves show $k T_{\rm eff,diss}$ for $f_{\rm c}=0.5$ and the green curves show $k T_{\rm eff,irr}$. At large $R$, both profiles are $\propto R^{-3/4}$ (the black lines).
}
\label{fig:T_R}
\end{figure}

We approximate the local spectrum of the quasi-thermal emission by a diluted blackbody,
\begin{equation}
B_{\nu}^{\rm db} = \kappa^{-4} B_{\nu}(\kappa T_{\rm eff}),
\end{equation}
where $B_{\nu}$ is the Planck function, $\kappa = T_{\rm col}/T_{\rm eff}$ is the spectral hardening factor and $T_{\rm col}$ is locally observed color temperature. For the angular distribution of local disk emission we use
\begin{equation}
I_{\nu}(\mu) = 3{1 + \delta_{\mu} \mu \over  3 + 2 \delta_{\mu}} B_{\nu}^{\rm db},
\end{equation}
where $I_{\nu}$ is the specific intensity in the disk rest frame, $\mu = \cos \theta$, $\theta$ is the angle with respect to the normal to the disk measured in the disk rest frame, $\delta_{\mu} = 0$ is for a locally isotropic emission and  $\delta_{\mu} = 2.06$ is for the classical scattering limit. $I_{\nu}$ is then convolved with the disk-to-observer transfer functions of the \texttt{reflkerr} model to find the observed spectrum of the quasi-thermal emission.

\begin{figure}
\centerline{\includegraphics[height=7.cm,angle=-90]{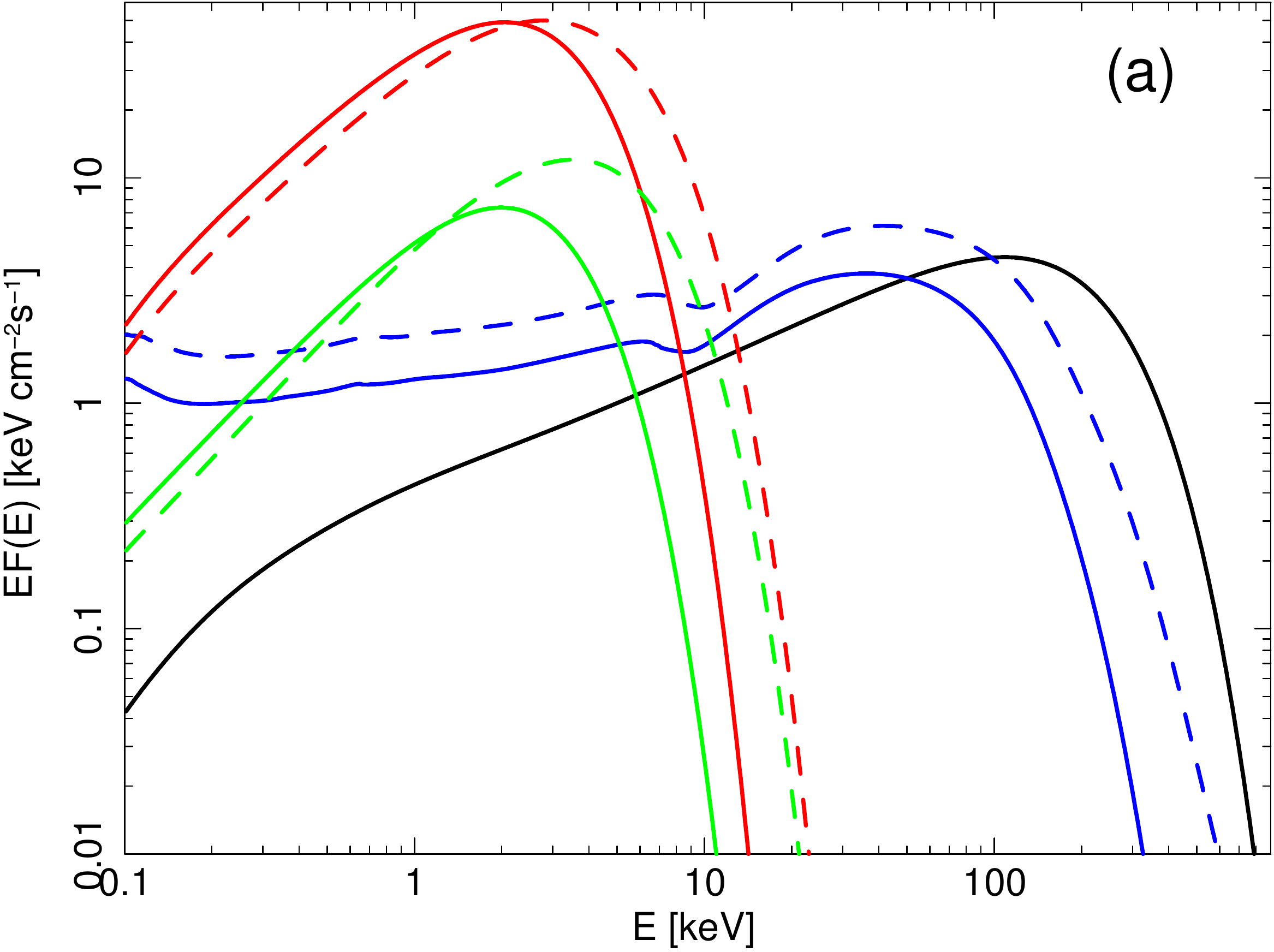}}
\centerline{\includegraphics[height=7.cm,angle=-90]{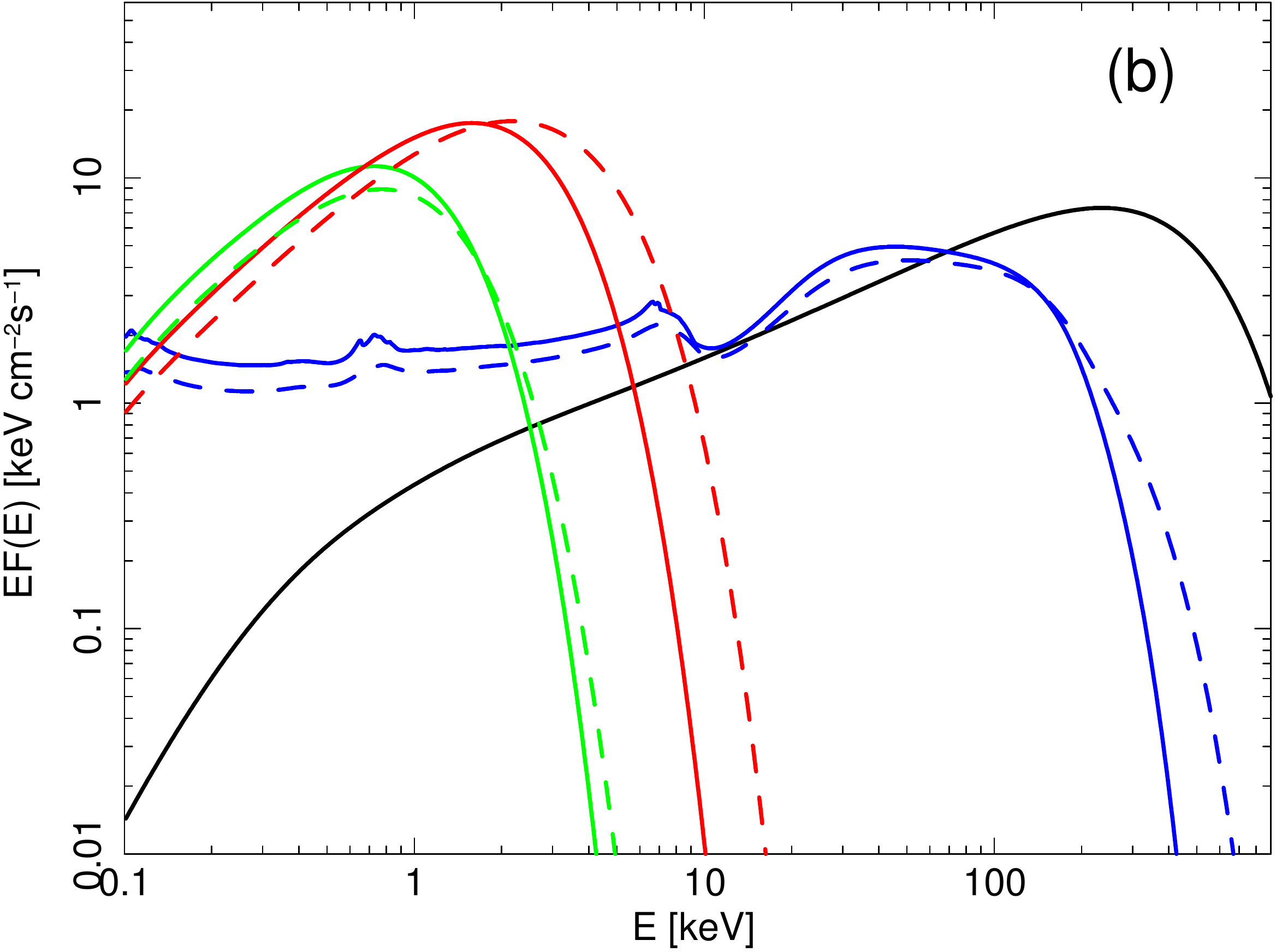}}
\caption{Spectral components in the lamppost geometry observed at $i = 9 \degr$ and $d = 3.5$ kpc, for $M = 9.6 \msun$, $R_{\rm in} = R_{\rm ISCO}$, $a_*=0.998$, $a = 0.5$, $\kappa = 1.7$, $\delta_{\mu} = 0$, $f_{\rm c} = 0.5$, $\log_{10} \xi = 3.5$, $kT_{\rm seed} = 0.1$\,keV, $kT_{\rm e} = 140$\,keV and (top) $L_{\rm l} = 1.6 \times 10^{38}$\,erg\,s$^{-1}$, $H =2 R_{\rm g}$ and (bottom) $L_{\rm l} = 5.7 \times 10^{37}$\,erg\,s$^{-1}$, $H=30 R_{\rm g}$. The corresponding disk temperature profiles are shown in Figure \ref{fig:T_R} above.
Thermal Comptonization is shown in black, reflection in blue, the diluted blackbody emission due to irradiation in green and due to internal dissipation in red. The last two components are shown separately for illustration only; the actual emission would form a single diluted blackbody. The solid curves show the spectral components observed at $i = 9 \degr$ and the dashed curves show spectra for $i = 45 \degr$. The Comptonization components are the same for both $i$.
}
\label{fig:spectra}
\end{figure}

\begin{figure}
\centerline{\includegraphics[height=7.cm,angle=-90]{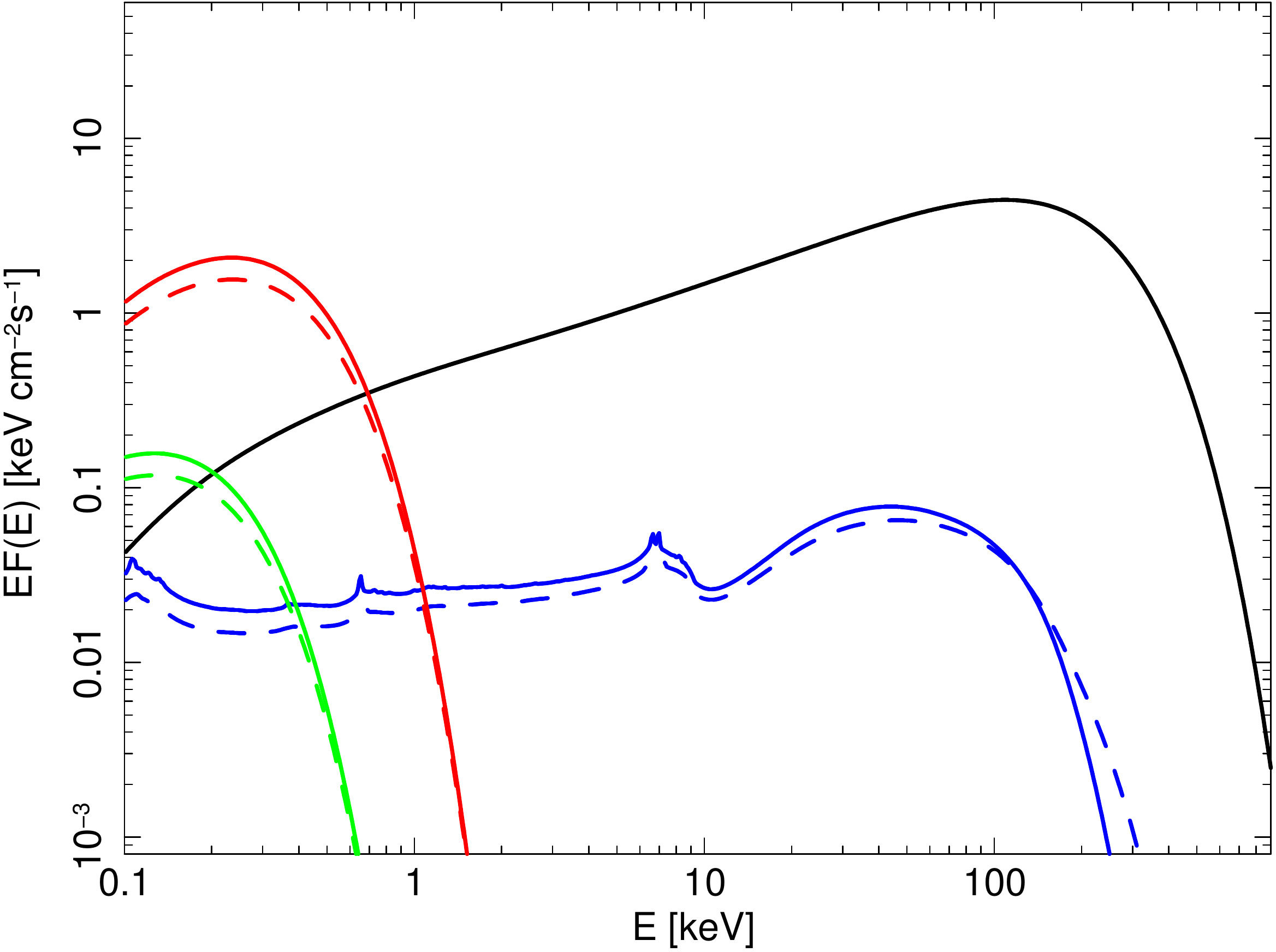}}
\caption{Similar to  Figure \ref{fig:spectra} (top) but for $R_{\rm in}=100 R_{\rm g}$, $\delta=0$.
}
\label{fig:h2_100_0}
\end{figure}

\begin{figure}
\centerline{\includegraphics[height=7.cm,angle=-90]{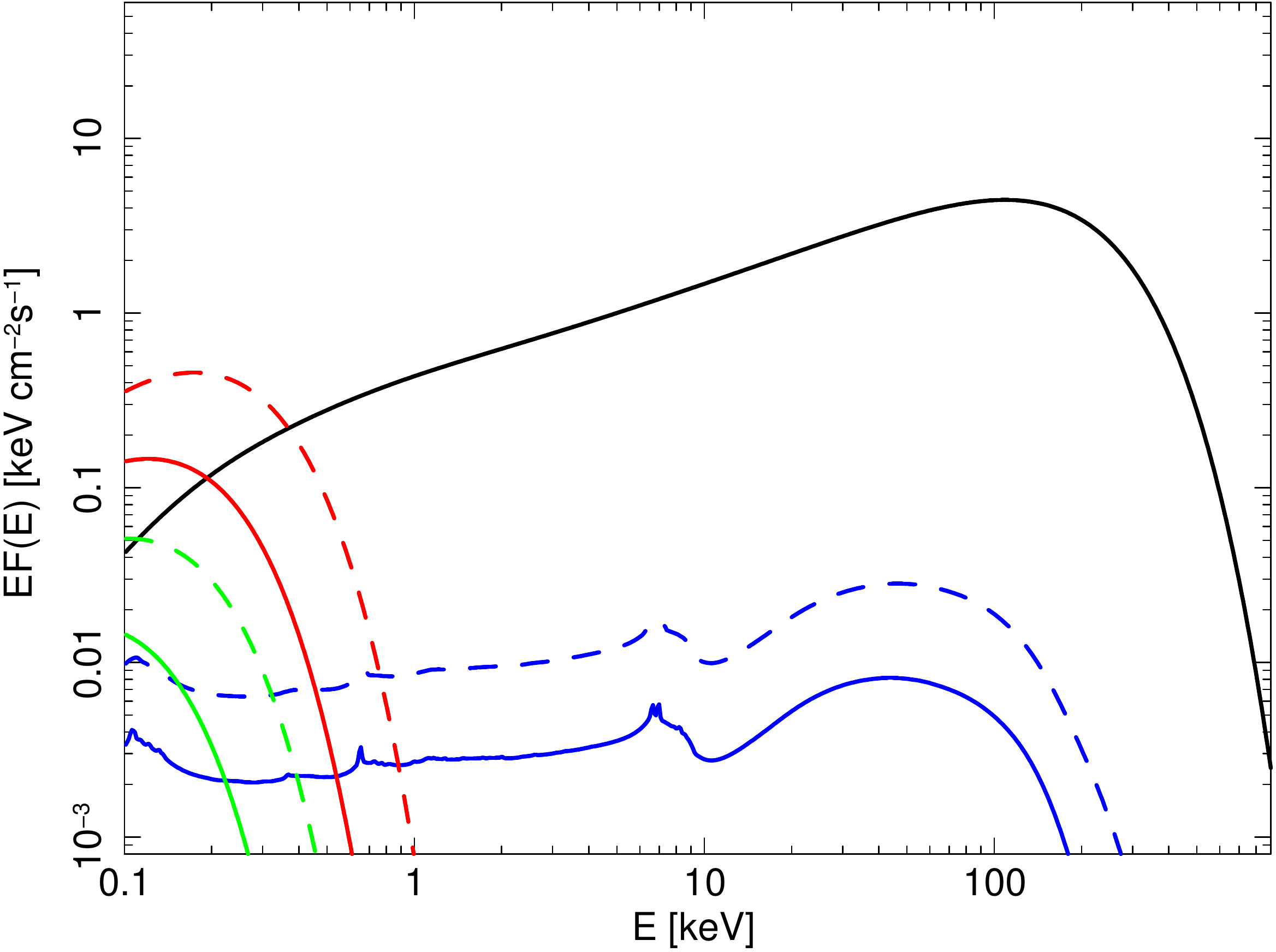}}
\caption{Similar to  Figure \ref{fig:spectra} (top) but for $R_{\rm in}=100 R_{\rm g}$, $\delta=1$, $L_{\rm l}=1.1 \times 10^{37}$\,erg\,s$^{-1}$ for $i=9 \degr$ and $L_{\rm l}=4.7 \times 10^{37}$\,erg\,s$^{-1}$ for $i=45 \degr$. Contribution to the direct component from the bottom lamp is stronger than that from the top lamp by a factor of 13 for $i = 9 \degr$ and by a factor 2.4 for $i = 45 \degr$, which reduces $L_{\rm l}$ compared to models with either $R_{\rm in} = R_{\rm ISCO}$ or $\delta=0$.
}
\label{fig:h2_100_1}
\end{figure}

Example temperature profiles are shown in Figure \ref{fig:T_R} and the corresponding observed spectra for a model with an untruncated disk are shown in Figure \ref{fig:spectra}. Note that $T_{\rm diss}(R)$ strongly differs from $T_{\rm irr}(R)$ at $R \lesssim 10 R_{\rm g}$. As a result, the relative contributions of the quasi-thermal emissions due to irradiation and due to internal dissipation will strongly depend on the inclination, $i$, as can be seen in Figure \ref{fig:spectra}.  

Figures \ref{fig:h2_100_0} and \ref{fig:h2_100_1} show spectra for the disk truncated at $R_{\rm in} = 100 R_{\rm g}$ for $\delta=0$ and 1, which allows us to illustrate the effect of the bottom lamp. Here $\delta$ is the attenuation factor of the bottom lamp; $\delta=1$ means that the region at $R < R_{\rm in}$ is fully transparent, for $\delta=0$ the bottom lamp is fully obscured, see also fig.\ A2 in \citet{Niedzwiecki19}. For $\delta=1$, the direct Comptonization component is enhanced by the (gravitationally focused) emission of the bottom lamp, while the reflection as well as the thermalized re-emission remain almost the same as for $\delta=0$. However, this is seen as a decrease of the amplitudes of the latter two components, as in all our plots the primary component has the same normalization at 1 keV. This effect is angle-dependent, because the contribution from the bottom lamp strongly depends on $i$.

We then compare the features of our model with those of the high-density version of the {\tt reflionx} code \citep{Ross99, Ross07, Tomsick18}, to which a relativistic convolution model of the {\tt relconv} family \citep{Dauser10, Dauser13} can be applied. The {\tt reflionx} model gives the reflection and reprocessing spectrum for a single value of the illuminating flux, which follows from the fitted values of $\xi$ and $n$. It can also include a contribution from a single value intrinsic dissipation. Thus, this spectrum consists of a quasi-thermal component corresponding to a single value of the effective temperature, while our model takes account of the distribution of $T_{\rm eff}$ in the disk. On the other hand, {\tt reflionx} spectra are obtained in a self-consistent way, though more approximately than the spectra from the {\tt xillver} models (which, however, can be used at this time only up to $n=10^{19}$\,cm$^{-3}$). This is a major advantage of the high-density {\tt reflionx} model. These spectra are then convolved with relativistic effects for a given geometry, either assuming a phenomenological irradiation profile approximating a corona or a lamppost. Thus, the range of the radius from which the reflection/reprocessing takes place is taken into account only in the relativistic effects, but not in the intrinsic emission.

Finally, we comment on our parametrization of the lamppost power. It assumes that the total accretion efficiency is that of the standard GR disk models, and that efficiency is divided into powering the intrinsic disk emission and the lamppost. Therefore, the actual disk emission is weaker than that of the standard model \citep{NT73}. On the other hand, models of jets usually assume the jet power, $P_{\rm j}$, to be independent of the disk dissipation and express it in units of $\dot M c^2$, where $\dot M$ is obtained from the observed disk emission, usually assuming a large efficiency, e.g., $\epsilon= 0.4$ \citep{Zamaninasab14}. The distribution of $P_{\rm j}/\dot M c^2$ estimated using different methods based on observations exceeds unity, e.g., \citet{Pjanka17}. However, the actual $\dot M$ can be underestimated in this procedure if a part of the power from the disk dissipation is transferred to the jet and/or the disk is truncated, in which cases the accretion efficiency will be lower.

\bibliography{allbib}{}
\bibliographystyle{aasjournal}

\label{lastpage}
\end{document}